\documentclass[prb,twocolumn,preprintnumbers,amsmath,amssymb]{revtex4}

\usepackage{graphicx}
\usepackage{dcolumn}
\usepackage{bm}
\usepackage{float}

\begin{document}

\preprint{APS/123-QED}

\title{Crystal nucleation of hard spheres using molecular dynamics, umbrella sampling and forward flux sampling: A comparison of simulation techniques}

\author{L. Filion}
 \affiliation{Soft Condensed Matter, Debye Institute for NanoMaterials Science, Utrecht University,
  Princetonplein 5, 3584 CC Utrecht,
  The Netherlands}
\author{M. Hermes}
 \affiliation{Soft Condensed Matter, Debye Institute for NanoMaterials Science, Utrecht University,
  Princetonplein 5, 3584 CC Utrecht,
  The Netherlands}
\author{R. Ni}
 \affiliation{Soft Condensed Matter, Debye Institute for NanoMaterials Science, Utrecht University,
  Princetonplein 5, 3584 CC Utrecht,
  The Netherlands}
\author{M. Dijkstra}
 \affiliation{Soft Condensed Matter, Debye Institute for NanoMaterials Science, Utrecht University,
  Princetonplein 5, 3584 CC Utrecht,
  The Netherlands}
\date{\today}

\begin{abstract}
Over the last number of years several simulation methods have been introduced to  study 
rare events such as nucleation.  In this paper we examine the crystal nucleation rate of hard spheres 
using three such numerical techniques: molecular dynamics, forward flux sampling and a Bennett-Chandler type 
theory where the nucleation barrier is determined using umbrella sampling simulations.  The resulting 
nucleation rates are compared with the experimental rates of Harland and Van Megen
[J. L. Harland and W. van Megen, Phys. Rev. E {\bf 55}, 3054 (1997)], 
Sinn {\it et al.} [C. Sinn {\it et al.}, Prog. Colloid Polym. Sci. {\bf 118}, 266 (2001)]
 and Sch\"atzel and  Ackerson [K. Sch\"atzel and  B.J. Ackerson, Phys. Rev. E, {\bf 48}, 3766 (1993)]
 and the 
predicted rates for monodisperse and 5\% polydisperse hard spheres of  Auer and  Frenkel
[S. Auer and D. Frenkel, Nature {\bf 409}, 1020 (2001)].
When the rates are examined in long-time diffusion units, we find agreement between all the theoretically 
predicted nucleation rates, however, the experimental results display a markedly different behaviour for low supersaturation.
Additionally, we examined the pre-critical nuclei arising in the molecular dynamics, forward flux sampling, and umbrella 
sampling simulations.  The structure of the nuclei appear independent of the simulation method, and in all cases, 
the nuclei contain on average significantly more face-centered-cubic ordered particles than hexagonal-close-packed ordered particles.
\end{abstract}


\maketitle


\section{Introduction}
\label{sec:intro}

Nucleation processes are ubiquitous in
both natural and artificially-synthesized systems.  
However, the occurrence of a nucleation event is often rare
and difficult to examine both experimentally and theoretically.  

Colloidal systems are almost ideal model systems for
studying nucleation phenomena.  Nucleation and the proceeding
crystallization in such systems often take place on
experimentally accessible time scales, and 
due to the size of the particles, they are accessible to 
a wide variety of scattering and imaging techniques, such as (confocal) 
microscopy, \cite{dinsmore_three-dimensional_2001}
holography, \cite{lee_characterizing_2007} and light and x-ray scattering.  Additionally, 
progress in particle synthesis, \cite{glotzer_anisotropy_2007} solvent manipulation, and the 
application of external fields \cite{yethiraj_colloidal_2003}  allows for significant 
control over the interparticle interactions, allowing 
for the study of a large variety of nucleation processes.

One such colloidal system is the 
experimental realization of ``hard'' spheres comprised of sterically 
stabilized polymethylmethacrylate (PMMA)   particles
suspended in a liquid mixture of decaline and carbon disulfide. \cite{harland_crystallization_1997}
Experimentally, the phase behaviour of 
such a system has been examined by Pusey and Van Megen \cite{pusey_phase_1986} and maps
well onto the phase behaviour predicted for hard spheres. 
Specifically when the effective volume fraction of their system is scaled to reproduce the 
freezing volume fraction of hard spheres ($\eta = 0.495$)
the resulting melting volume fraction is $\eta = 0.545 \pm 0.003$ \cite{pusey_phase_1986}
which is in good agreement with that 
predicted for hard spheres. \cite{hoover_melting_1968} The 
nucleation rates have been measured using light scattering by  Harland and 
Van Megen, \cite{harland_crystallization_1997}
 Sinn {\it et al.}, \cite{sinn_solidification_2001}  Sch\"atzel and  Ackerson \cite{schatzel_density_1993} 
and predicted theoretically by  Auer and  Frenkel. \cite{auer_prediction_2001}

On the theoretical side, hard-sphere systems are one of the simplest 
systems which can be applied to the study 
of colloidal and nanoparticle systems, and generally, 
towards the nucleation process itself. As such, 
it is an ideal system to examine various computational methods 
for studying nucleation, and comparing the results 
with experimental data.  
Such methods include, but are not limited to, molecular dynamics (MD) simulations, umbrella 
sampling (US), forward flux sampling (FFS), and transition path sampling (TPS).
It is worth noting here that Auer and Frenkel \cite{auer_prediction_2001} used umbrella sampling
simulations to study crystal nucleation of hard spheres and found a significant difference between their predicted rates and 
the experimental rates of Refs. \onlinecite{harland_crystallization_1997,sinn_solidification_2001,schatzel_density_1993}.  However, it was unclear
where this difference originated.  
In this paper we compare the nucleation rates 
for the hard-sphere system from MD, US and FFS simulations
with the experimental results of Refs. \onlinecite{harland_crystallization_1997, sinn_solidification_2001,schatzel_density_1993}.
We demonstrate that the three simulation techniques are consistent in their prediction of the nucleation
rates, dispite the fact that they treat the dynamics differently. Thus we conclude that the difference between 
the experimental and theoretical nucleation rates identified by Auer and Frenkel is not due to the simulation method.

A nucleation event occurs when a statistical fluctuation in a 
supersaturated liquid results in the formation of a crystal nucleus
large enough to grow out and continue crystallizing the surrounding fluid. 
In general, small crystal nuclei are continuously being formed
and melting back in a liquid. However, while most of these small nuclei will quickly melt, 
in a supersaturated liquid a fraction of these nuclei will 
grow out. Classical nucleation theory is the simplest theory available for describing this
process.  In CNT it is assumed that the free energy for making a small 
nuclei is given by a surface free energy cost which is proportional to the 
surface area of the nucleus and a bulk free energy gain proportional to its
 volume. More specifically, according to CNT 
the Gibbs free energy difference between a homogeneous bulk fluid and 
a system containing a spherical nucleus of radius $R$ is given by
\begin{equation}
\Delta G(R) = 4 \pi \gamma R^{2} -\frac{4}{3} \pi \left|  \Delta \mu \right|  \rho_{s} R^{3}
\end{equation}
where $\left| \Delta \mu \right| $ is the difference in chemical potential between the fluid and solid phases,
$\rho_{s}$ is the density of the solid, and $\gamma$ is the surface tension of the
fluid-solid interface. This free energy difference is usually referred to as the
nucleation barrier.  From this expression, the radius of the critical cluster is found to be
$R^{*} = 2 \gamma/\left| \Delta \mu\right| \rho_{s}$ 
and the barrier height is
$ \Delta G^{*} = 16 \pi \gamma^{3} / 3 \rho_{s}^{2} \left| \Delta \mu \right|^{2}$. 
Note that there is no system size dependence in CNT.

Umbrella sampling \cite{torrie_monte_1974,van_duijneveldt_computer_1992} 
is a method to examine the nucleation
process from which the nucleation barrier is
easily obtained.  The predicted barrier can then be used in combination 
with kinetic Monte Carlo (KMC) or MD simulations to determine the
nucleation rate. \cite{auer_prediction_2001}  In 
US an order parameter for the system is chosen and 
configuration averages for sequential values of the order parameter are taken.
In order to facilitate such averaging, the system 
is biased towards particular regions in configuration space.
The success of the method is expected to depend largely on
the choice of order parameter and biasing potential.
Note that the free energy barrier is only defined in equilibrium, and 
thus is only applicable to systems which are in (quasi-) equilibrium.


Forward flux sampling \cite{allen_sampling_2005,allen_simulating_2006,allen_forward_2006}
is a method of studying rare events, such as nucleation, in both 
equilibrium and non-equilibrium systems.  Using FFS, the transition rate constants (eg. the nucleation rate) for 
rare events can be determined when   
brute force simulations are difficult or even not possible.
In FFS, a reaction coordinate  $Q$ (similar to the order parameter in US) is introduced which 
follows the rare event. The transition rate between phase A and B is then 
expressed as a product of the flux 
($\Phi_{A\lambda_{0}}$) 
of trajectories crossing the A state boundary, typically denoted $\lambda_0$, and the 
probability ($P(\lambda_{B}|\lambda_{0})$) 
that a trajectory which has crossed this boundary will reach state B before returning to state A.
Thus the transition rate constant is written as
\begin{equation}
k_{AB} = \Phi_{A\lambda_0} P(\lambda_B|\lambda_0).
\end{equation}
Forward flux sampling facilitates the calculation of probability $ P(\lambda_B|\lambda_0)$  by breaking 
it up into a set of probabilities between sequential values of the reaction coordinate.
Little information regarding the details of the nucleation process is 
required in advance, and the choice of reaction coordinate is expected to be less important 
than the order parameter in US. Additionally, unlike US, FFS utilizes 
dynamical simulations and hence this technique does not assume that the
system is in (quasi-)equilibrium.

Molecular dynamics  and Brownian dynamics (BD) simulations are ideal for studying the time
evolution of systems, and, when possible, they are the natural technique
to study dynamical processes such as nucleation. Unfortunately, however, available
computational time often limits the types of systems
which can be effectively studied by these dynamical techniques. Brownian dynamics simulations,
which would be the natural choice to use for colloidal systems,
are very slow due to the small time steps required to handle the steep potential used
to approximate the hard-sphere potential. Event driven
MD simulations are much more efficient to simulate hard spheres and enable
us to study spontaneous nucleation of hard-sphere mixtures over a range of volume
fractions. The main difference between the two simulation methods 
regards how they treat the short-time motion of the particles.
Fortunately, the nucleation rate is only 
dependent on the long-time dynamics which are not sensitive to the details of the short-time 
dynamics of the system. \cite{pusey_hard_2009}

\begin{table}
\begin{tabular}{ccc}
\hline 
\hline 
$\eta$ & $\beta p \sigma^{3}$ & $\left| \Delta \mu \right|$ \ \\
\hline 
0.5214  & 15.0 & 0.34 \\
0.5284  & 16.0 & 0.44 \\
0.5316  & 16.4 & 0.48 \\
0.5348  & 16.9 & 0.53 \\
0.5352  & 17.0 & 0.54 \\
0.5381  & 17.5 & 0.58 \\
0.5414  & 18.0 & 0.63 \\
0.5478  & 19.1 & 0.74 \\
0.5572  & 20.8 & 0.90 \\
\hline
\end{tabular}
\caption{Packing fraction ($\eta=\pi\sigma^{3}N/6V$) , reduced pressure ($\beta p \sigma^{3}$)  and chemical 
potential difference between the fluid and solid phases( $\left| \Delta \mu \right|$) of the state points studied in this paper.
The chemical potential difference was determined using thermodynamic integration, \cite{frenkel_understanding_1996} and the equations of 
state for the fluid and solid are from Refs. \onlinecite{speedy_pressure_1997,speedy_pressure_1998} respectively. 
\label{tab:sims}}
\end{table}

In this paper we study in detail the application of US and FFS 
techniques to crystal nucleation of hard spheres, and predict the 
associated nucleation rates.
Combining these  nucleation rates  with results from MD simulations, 
we make predictions for the nucleation rates over a wide 
range of packing fractions  $\eta=0.5214-0.5572$, with corresponding 
pressures and supersaturations shown in Table \ref{tab:sims}.
We compare these theoretical nucleation rates  with the rates measured  
experimentally by Refs. \onlinecite{harland_crystallization_1997,sinn_solidification_2001,schatzel_density_1993}.

This paper is organized as follows: in section  \ref{sec:order}
we describe and examine the order parameter
used to distinguish between solid- and fluid-like particles throughout this paper, 
in section \ref{sec:MD}  we calculate 
essentially the ``exact'' nucleation rates using MD simulations, 
in sections \ref{sec:US}  and \ref{sec:FFS} we calculate the nucleation rates of hard spheres
using US and FFS respectively, and discuss difficulties in the 
application of these techniques, 
in section \ref{sec:results}  we summarize the theoretical results and compare the
predicted nucleation rates with the measured experimental rates of   Harland and
Van Megen, \cite{harland_crystallization_1997}  Sinn {\it et al.}, \cite{sinn_solidification_2001}
and  Sch\"atzel and  Ackerson \cite{schatzel_density_1993} and section \ref{sec:conclusions}
contains our conclusions.


\section{Order Parameter}
\label{sec:order}

In this paper, an order parameter is used 
to differentiate between liquid-like and solid-like particles and a cluster algorithm is 
used to identify the solid clusters.  For this study we have chosen to use
the local bond-order parameter introduced by
Ten Wolde {\it et al.}  \cite{wolde_simulation_1996,pieter_rein_ten_wolde_numerical_1998} in the study of 
crystal nucleation in a Lennard-Jones system.  This order parameter has been used in many crystal nucleation studies, 
including a previous study of hard-sphere nucleation by Auer and Frenkel. \cite{auer_prediction_2001} 

In the calculation of the local bond order parameter a
list of ``neighbours'' is determined for each particle.  
The neighbours of particle $i$ include all particles within 
a radial distance $r_{c}$ of particle $i$, and
the total number of neighbours is denoted $N_{b}(i)$. 
A bond orientational order parameter $q_{l,m}(i)$ for each particle is then defined as
\begin{equation}
q_{l,m}(i) = {\frac{1}{N_{b}(i)} \sum_{j=1}^{N_{b}(i)} Y_{l,m}(\theta_{i,j},\phi_{i,j})}
\end{equation}
where $Y_{l,m}(\theta,\phi)$ are the spherical harmonics, $m\in[-l,l]$ and $\theta_{i,j}$ and $\phi_{i,j}$ are the polar and azimuthal angles
of the center-of-mass distance vector $\mathbf{r}_{ij}=\mathbf{r}_{j}-\mathbf{r}_{i}$
with $\mathbf{r}_{i}$ the position vector of particle $i$.
Solid-like particles are identified as particles 
for which the number of connections per particle $\xi (i)$ is 
at least $\xi_{c}$ and 
where 
\begin{equation}
\xi (i) = \sum_{j=1}^{N_{b}(i)} H (d_{l}(i,j) - d_{c}),
\end{equation}
$H$ is the Heaviside step function, $d_c$ is the dot-product cutoff, and 
\begin{equation}
d_{l}(i,j) = \frac{\displaystyle \sum_{m=-l}^{l} {q}_{l,m}(i) {q}^{*}_{l,m}(j)}{\displaystyle {\left(\sum_{m=-l}^{l} |{q}_{l,m}(i)|^{2}\right)^{1/2}}{\left(\sum_{m=-l}^{l} |{q}_{l,m}(j)|^{2}\right)^{1/2}}}.
\end{equation}
A cluster contains all solid-like particles which have a 
solid-like neighbour in the same cluster.  Thus each particle can be
a member of only a single cluster. 

The parameters contained in this algorithm include the neighbour 
cutoff $r_{c}$, the dot-product cutoff $d_{c}$, the 
critical value for the number of solid-like neighbours $\xi_{c}$, and the 
symmetry index for the bond orientational order parameter $l$.
The solid nucleus of a hard-sphere crystal is expected
to have random hexagonal order, thus the symmetry index is 
chosen to be 6 in all cases in this study.  

To investigate the effect of the choice of $\xi_{c}$, we examined the number of 
correlated bonds per particle at the liquid-solid interface. To this end,  
we constructed a configuration in the coexistence 
region in an elongated box by attaching a box containing an equilibrated random-hexagonal-close-packed 
(RHCP) crystal to a box containing an equilibrated
fluid. Note that the RHCP crystal was placed in the box such that the hexagonal
layers were parallel to the interface. 
The new box was then equilibrated in an NPT MC simulation.  We
then examined the density profile of solid-like particles 
as determined by our order parameter using $r_{c}=1.4$, $d_{c}=0.7$ and $\xi_{c}=5,7$ and 9.  
As shown in Fig. \ref{fig:fluidsolid}, for all values of $\xi_{c}$ that we examined
the order parameter appears to consistently identify the  particles belonging to the bulk fluid and 
solid regions. For comparison we also show a typical configuration of the RHCP crystal in coexistance 
with the fluid phase.  The solid-like particles as defined by the order parameter are labelled 
according to the number of solid-like neighbours while the fluid-like particles are denoted by dots.
The main difference between these order parameters relates to distinguishing between
fluid- and solid-like  particles at the fluid-solid interface. Unsurprisingly, the location 
of the interface seems to shift in the direction of the bulk solid as $\xi_{c}$ is increased. We note that the dips in the density profile 
correspond to HCP stacked layers which are more pronounced for higher values of $\xi_{c}$. 

\begin{figure}[h]
\centering
\includegraphics[width=0.45\textwidth]{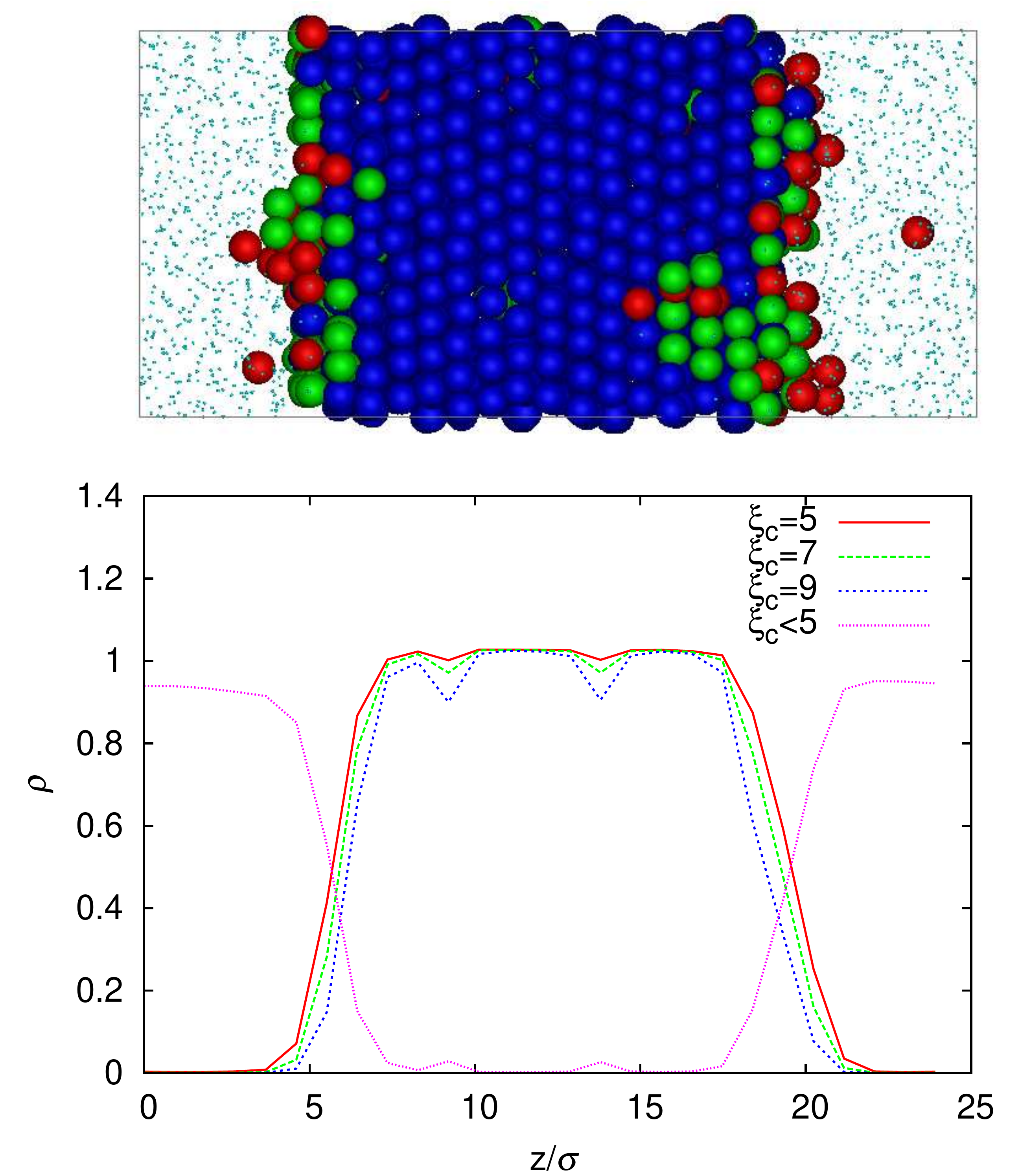}
\label{fig:fluidsolid}
\caption{\label{fig:fluidsolid} 
Top:
A typical configuration  
of an equilibrated random-hexagonal-close-packed 
(RHCP) crystal in coexistance with an equilibrated
fluid.  The crystalline particles are labelled according to three different 
crystallinity criteria: the red particle have between  $\xi=5$ and 6 crystalline bonds, 
the green particles have between $\xi=7$ and 8 crystalline bonds and the blue particles have 
$\xi \ge 9$ or more crystalline bonds.  The fluid-like particles ($\xi<5$) are 
denoted by dots.
Bottom:  The density profile of particles with a minimum number of neighbours $\xi$ as labelled.  
Note that the dips in the density profile correspond to HCP stacked layers. }
\end{figure}

\section{Molecular Dynamics}
\label{sec:MD}
\subsection{Nucleation Rates}
In MD simulations the equations of motion are
integrated to follow the time evolution of the system. Since the hard-sphere potential is
discontinuous the interactions only take place when particles collide.
Thus the particles move in straight lines (ballistic) until they
encounter another particle with which they perform an elastic collision. \cite{alder_studies_1959} These
collision events are identified and handled in order of occurrence using an event driven simulation.

In theory, using an MD simulation to determine nucleation rates is quite simple.
Starting with an equilibrated fluid configuration, an MD simulation is used to evolve the system
until the largest cluster in the system exceeds the critical nucleus size. The MD time associated with
such an event is then measured and averaged over many initial configurations.  The nucleation rate is 
given by 
\begin{equation}
 k = \frac{1}{\left<t\right> V}
\end{equation}
where $V$ is the volume of the system and $\left<t\right>$ is the average time
to form a critical nucleus.
Measuring this time is relatively easy for low
supersaturations where the nucleation times are relatively long compared to the nucleation
event itself, which corresponds with a steep increase in the crystalline fraction of the system.
However, for high supersaturations pinpointing the time of a nucleation event
is more difficult. Often many nuclei form immediately and the critical nucleus sizes must be estimated from 
 CNT or US simulations. Additionally,  the precise details of the initial
configuration can play a role at high supersaturations since
the equilibration time of the fluid is of the same order of magnitude as the nucleation
time.

For the results in this paper, we performed MD simulations with up to 100,000 particles in
a cubic box with periodic boundary conditions in an NVE ensemble. Time was measured
in MD units $\sigma\sqrt{m/k_BT}$. The order
parameter was measured every 10 time units and when the largest cluster exceeded the
critical size by 100 percent we estimated the time $\tau_\mathrm{nucl}$ at which the
critical nucleus was formed using stored previous configurations.
We performed up to 20 runs for every density and averaged the nucleation times.

\begin{table}[!hbt]

\begin{tabular}{l | l | l}
\hline
\hline
Volume fraction & Average nucleation time & Rate \\
$\eta$ & $t\sqrt{k_BT/(m\sigma^2)}$ & $k \sigma^5/ (6D_l)$ \\
\hline
0.5316  & $1\cdot10^6$   & 5$\cdot 10^{-9}$ \\
0.5348  & $1.7\cdot10^4$ & 3.6$\cdot 10^{-7}$ \\
0.5381  & $1.4\cdot10^3$ & 5.3$\cdot 10^{-6}$ \\
0.5414  & $2.0\cdot10^2$ & 4.3$\cdot 10^{-5}$ \\
0.5478  & 42         & 3.0$\cdot 10^{-4}$ \\
0.5572  & 10         & 2.4$\cdot 10^{-3}$ \\
\hline
\end{tabular}
\caption{The average nucleation time, obtained from MD simulations, to form a
critical cluster that grew out and filled the box. The last column contains the rate (k)
in units of $(6D_l)/\sigma^{5}$.
\label{table_mdrate}}
\end{table}

The results are shown in Table \ref{table_mdrate}. The nucleation times shown here are for a system
of $2.0\cdot 10^4$ particles and in MD time units. To compare with other data we convert
the MD time units to units of $\sigma^{2}/(6D_l)$ with $D_l$ the long-time diffusion coefficient
measured in the same MD simulations.
We were not able to measure the long-time diffusion coefficients for high densities
because our measurements were influenced by crystallization.
We used the fit obtained by Zaccarelli {\it et al.}~\cite{zaccarelli_crystallization_2009} who used polydisperse
particles to prevent crystallization. For $\eta<0.54$, we find good agreement between our data for $D_{L}$ and this fit.


\section{Umbrella Sampling}
\label{sec:US}

\subsection{Gibbs Free-Energy Barriers}
\label{sec:gibbsbarriers}
Umbrella sampling is a technique developed by Torrie and Valleau to study systems where Boltzmann-weighted
sampling is inefficient. \cite{torrie_monte_1974}  
This method has been applied frequently to study rare events, such as nucleation, 
\cite{van_duijneveldt_computer_1992} and specifically has been applied 
in the past to study the nucleation of hard spheres. \cite{auer_prediction_2001} In general, umbrella sampling is 
used to examine parts of configurational space which are unaccessible by traditional schemes, eg. Metropolis Monte
Carlo simulations.  Typically, a biasing potential is added to the true interaction potential
causing the system to oversample a region of configuration space.  The biasing potential, however,
is added in a manner such that is it easy to ``un''-bias the measurables. 

In the case of nucleation, while it is simple to sample the
fluid, crystalline clusters of larger sizes will be rare, and as such, impossible to sample on reasonable time scales.  
The typical biasing potential for studying nucleation is given by \cite{ten_wolde_numerical_1996,wolde_simulation_1996}
\begin{equation}
U_{\mathrm{bias}}(n(\mathbf{r}^N)) = \frac{\lambda}{2} (n(\mathbf{r}^N) - n_{C})^{2}
\end{equation}
where $\lambda$ is a coupling parameter, $n(\mathbf{r}^N)$ is the size of the largest cluster  
associated with configuration $\mathbf{r}^{N}$, and $n_{C}$ is the 
targeted cluster size.  By choosing $\lambda$ carefully, the simulation will fluctuate around the 
part of configurational space with $n(\mathbf{r}^N)$ in the vicinity of $n_{C}$. The expectation value of an observable $A$ 
is then given by
\begin{equation}
\left<A\right> = \frac{\left<A/W(n(\mathbf{r}^N))\right>_{\mathrm{bias}}}{\left<1/W(n(\mathbf{r}^N))\right>_{\mathrm{bias}}}
\end{equation}
where 
\begin{equation}
W(x) = e^{-\beta U_{\mathrm{bias}}(x)}.
\end{equation}
Using this scheme to measure the probability distribution $P(n)$ for clusters of size $n$, 
the Gibbs free energy barrier can be determined by \cite{auer_numerical_2004}
\begin{equation}
\beta \Delta G (n) = \mathrm{constant} - \ln(P(n)).
\end{equation}
Many more details on this method are given elsewhere. \cite{auer_numerical_2004,frenkel_understanding_1996}

\begin{figure}[h]
\centering
\includegraphics[width=0.45\textwidth]{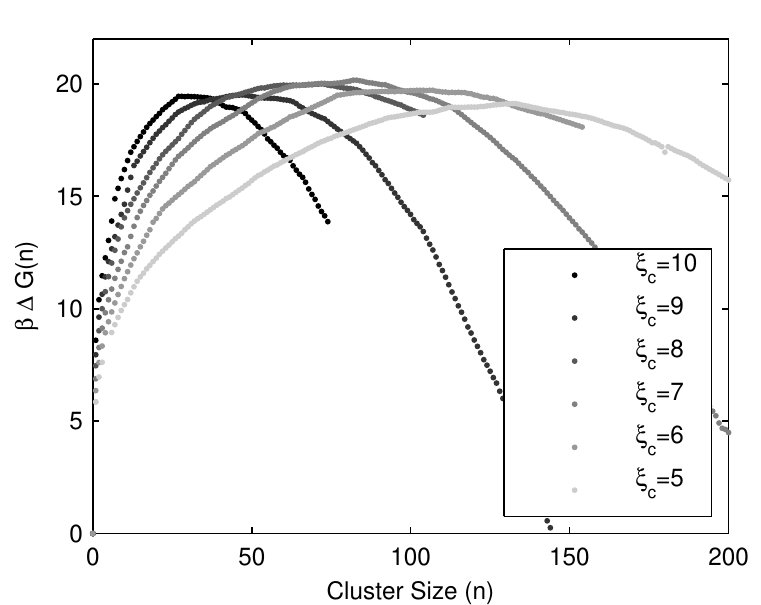}
\label{fig:simulationbarriers}
\caption{\label{fig:simulationbarriers} 
Gibbs free energy barriers $\beta \Delta G(n)$ as a function of cluster-size $n$ as obtained from umbrella sampling simulations 
at a reduced pressure of $\beta p \sigma^{3}=17$ for varying critical number of solid-like neighbours $\xi_{c}$ as labelled. 
For $\xi_c=5,7,9$, the neighbour cutoff is $r_c=1.4$ and for $\xi_c=6,8,10$, $r_c=1.3$. In all
cases the dot product cutoff is $d_c=0.7$. }
\end{figure}

\begin{figure}[h]
\centering
\includegraphics[width=0.45\textwidth]{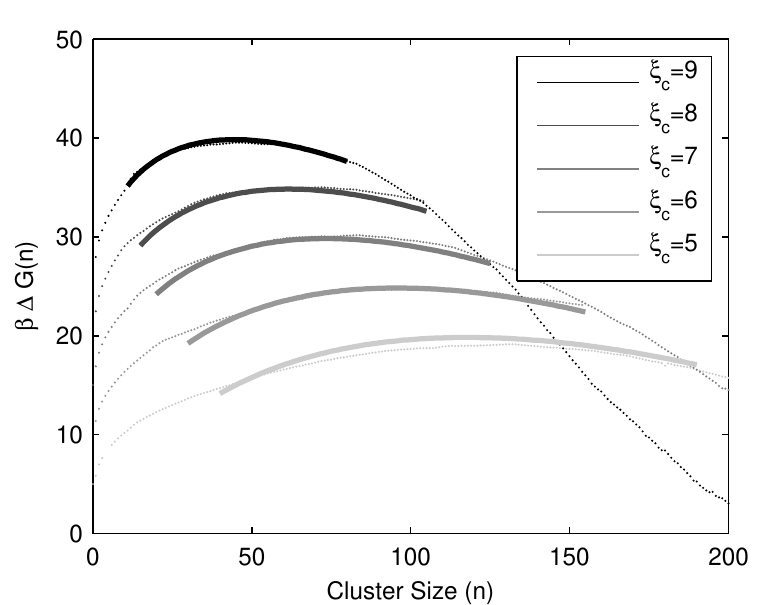}
\label{fig:CNTbarriers}
\caption{\label{fig:CNTbarriers}   
Classical nucleation theory fits (thick lines)  to the Gibbs free energy barriers obtained from umbrella sampling simulations 
at a reduced pressure of $\beta p \sigma^{3}=17$ for varying $\xi_{c}$ 
as labelled. Note that the CNT radius ($R_{CNT}$) is related to the  radius ($R(\xi_c)$) measured  by
umbrella sampling by $R(\xi_c) = R_{\mathrm{CNT}} + \alpha(\xi_{c})$, where $\alpha(\xi_{c})$ is a constant that corrects for the 
different ways the various order parameters identify the particles at the fluid-solid interface. The fit parameters
are given in Table \ref{table:barrierfits}.  We have shifted the barriers for $\xi_c=6-9$ by $5,10,15,20$ $k_BT$ respectively
for clarity}
\end{figure}

\begin{figure}[h]
\centering
\includegraphics[width=0.45\textwidth]{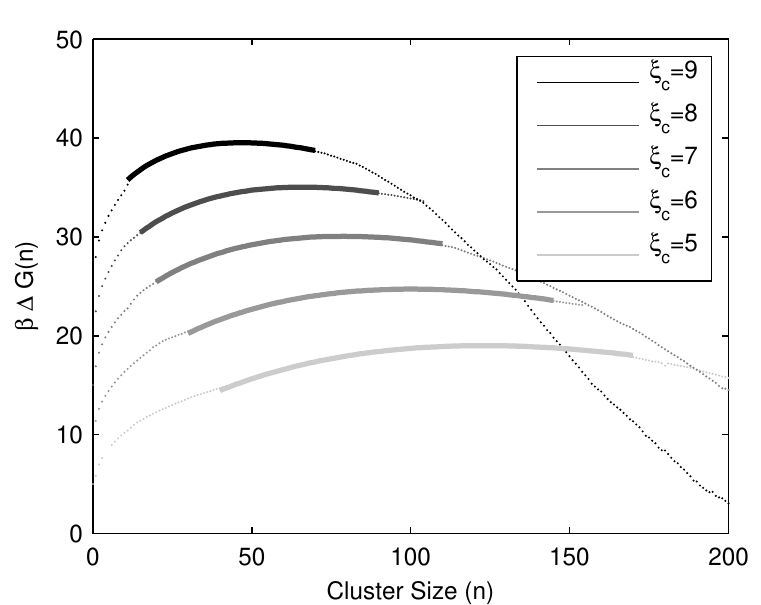}
\label{fig:ACNTbarriers}
\caption{\label{fig:ACNTbarriers} Fits of an adjusted classical nucleation theory (ACNT) 
presented in Section \ref{sec:gibbsbarriers}  to the Gibbs free energy barriers predicted using  umbrella sampling simulations 
at a reduced pressure of $\beta p \sigma^{3}=17$ and for varying $\xi_{c}$ 
as labelled. Note that the CNT radius ($R_{\mathrm{CNT}}$) is related to the  radius measured  by
umbrella sampling by $R(\xi_c) = R_{CNT} + \alpha(\xi_{c})$, where $\alpha(\xi_{c})$ is a constant. The fit parameters
are given in Table \ref{table:barrierfits}.  We have shifted the barriers for $\xi_c=6-9$ by $5,10,15,20$ $k_BT$ respectively 
for clarity.}
\end{figure}


\begin{table*}
\begin{tabular}{cccccccccccccc}
\hline
\hline
&  $ \beta \left| \Delta \mu \right| $ & $\beta \gamma \sigma^{2}$ & $R_{\mathrm{CNT}}^{*}$ & $\alpha(5)$ & $\alpha(6)$ & $\alpha(7)$ & $\alpha(8)$ & $\alpha(9)$ & $ c(5)$ & $c(6)$  & $c(7)$ &$c(8)$ &$c(9)$ \\ 
\hline
CNT &  0.54  & 0.76                      & 2.49          &-0.425      &-0.231      &-0.000       & 0.139       & 0.380       & & & & & \\
ACNT&  0.54  & 0.61                      & 2.01          &-0.961       &-0.765     &-0.551       &-0.402       &-0.148       & 8.75   & 9.46   & 9.81  & 9.78 & 9.28 \\
\hline
\end{tabular}
\label{table:barrierfits}
\caption{Numerical values for the parameters associated with the fits in Figs. \ref{fig:CNTbarriers} and \ref{fig:ACNTbarriers} 
for classical nucleation theory and the adjusted classical nucleation theory presented in this paper.}
\end{table*}

For a pressure of $\beta p \sigma^{3} = 17$, corresponding to a 
supersaturation of $\beta \left| \Delta \mu \right|  = 0.54$, we examine the effect of one of the order parameter variables, 
namely $\xi_{c}$,  on the prediction of the 
nucleation barriers. 
The barriers predicted by US using 
 $\xi_{c}=5,6,7,8,9$ and 10  are shown in Fig. \ref{fig:simulationbarriers}. Note that the height of the 
barriers does not depend on $\xi_c$ within error bars.
In general, for larger values of $\xi_{c}$ more particles are
identified as fluid as compared with smaller values of $\xi_{c}$.  This is
consistent with the differences between these order parameters as demonstrated in Fig. \ref{fig:fluidsolid}.

Taking the previous discussion on order parameters into consideration, we fit the barriers corresponding 
to $\xi_{c}=5,6,7,8$ and 9 using CNT where we assume there exists a CNT radius $R_{\mathrm{CNT}}$ which differs 
from the radius $R(\xi_c)$  measured by the order parameter. We assume that 
the difference ($\alpha$) is a constant for each value of the critical number of solid-like neighbours $\xi_c$ which
corrects for the different ways the various order parameters identify the particles at the fluid-solid interface:
\begin{equation}
R(\xi_c) = R_{\mathrm{CNT}} + \alpha(\xi_{c}).
\end{equation}
Note that we have assumed that the  cluster size $n$ can be related to the cluster radius $R(\xi_c)$ by
\begin{equation} 
n(\xi_c) = \frac{4 \pi R(\xi_c)^3 \rho_s}{3}.
\end{equation}
Fitting all barriers simultaneously for the surface tension, and the various $\alpha(\xi_{c})$, we obtain
the fits displayed in Fig. \ref{fig:CNTbarriers}. 
From the various values of $\alpha$, the associated critical CNT radius ($R_{\mathrm{CNT}}^{*}$) can be determined.
We find $R_{\mathrm{CNT}}^{*}=2.49\sigma$. Additionally, we find a surface tension of $\beta \gamma \sigma^{2}=0.76$
which roughly agrees with the results of  Auer and Frenkel who obtained 
surface tensions of $\beta \gamma \sigma^{2} =0.699, 0.738$ and $0.748$  for pressures
 $\beta p \sigma^{3} = 15, 16$ and 17 respectively. \cite{auer_prediction_2001} 
However, recent calculations by Davidchack {\it et al.}  \cite{davidchack_anisotropic_2006} of the surface tension at 
the fluid-solid coexistence find $\beta \gamma \sigma^{2} = 0.574, 0.557$ and 0.546 for the 
crystal planes (100), (110), and (111) respectively.
For a spherical nucleus, the surface tension is expected to be an average over the crystal planes.  Thus our
result for the surface tension and that of Ref. \onlinecite{auer_prediction_2001} appear to be an overestimate.

There have been a number of 
papers discussing possible corrections 
to CNT (eg. Refs. \onlinecite{ryu_validity_2010,ford_nucleation_1997}). 
Recent work on the 2d Ising model, a system 
where both the surface tension and supersaturation are known analytically, 
demonstrated that in order to match a nucleation barrier 
obtained from US to CNT, two correction terms were 
required, specifically a term proportional to $\log(N)$  as well as a constant
shift in $\Delta G$ which we define as $c$. \cite{ryu_validity_2010} The US barrier is only 
expected to match CNT near the top of the barrier where the $\log(N)$ term is
almost a constant. Thus, we propose fitting the barrier to an adjusted 
expression for CNT (ACNT), by adding a constant $c$ to Eq. 1. 
Fitting the US barriers with this 
proposed form for the Gibbs free energy barrier, where we assume $c$ is a function of $\xi_{c}$,  
we obtain the fits displayed in Fig. \ref{fig:ACNTbarriers}. In this case we find a surface tension 
$\beta \gamma \sigma^{2} =0.61$, and the values for $\alpha(\xi_{c})$ and $c(\xi_{c})$ are given in Table \ref{table:barrierfits}. 
The difference
in the various $c(\xi_{c})$ are around 1k$_{B}$T and correspond well to the difference in 
heights of the barriers.  More strikingly, the surface tension predicted 
from this proposed free energy barrier is in much better agreement with
recent calculations of Davidchack {\it et al.}, \cite{davidchack_anisotropic_2006}
than the surface tension we calculate using classical nucleation theory directly.
We would like to point out here that due to the simple form of the nucleation barrier,
it is difficult to be certain of any fit with more than one fitting parameter, as there 
are many combinations of parameters which fit almost equally well.

Using both expressions for the Gibbs free energy barrier, namely CNT and ACNT,
we were unable to fit the barrier corresponding to $\beta p \sigma^3=17$ and $n_{c}=10$ 
simultaneously with the other predicted  barriers
for the same pressure.
We speculate that our difficulty in fitting the barrier at $\xi_{c}=10$  
stems from an ``over-biasing'' of the system.
Specifically, by using $\xi_{c}=10$ the biasing potential could cause the system to
sample more frequently more ordered clusters, and hence change slightly the region of phase space 
available to the US simulations. In general, the least biased systems would be expected to 
explore the largest region of phase space resulting in the best results. 

In conclusion, with the exception of $\xi_{c}=10$, the value of $\xi_{c}$ 
used in the order parameter did not
appear to have an effect on the nucleation barriers once the difference in their 
measurements of the solid-liquid interface was taken into consideration.
Finally, for use in our nucleation rate calculations (section \ref{sec:usrates})  we also calculated
the Gibbs free energy $\Delta G(n)$ for reduced pressures $\beta p \sigma^{3}=15$ and 16 using 
umbrella sampling simulations.  We present the barrier heights in Table \ref{tab:USrates}.

\subsection{Umbrella Sampling Nucleation Rates}
\label{sec:usrates}
The nucleation barriers as obtained from US simulations can be used to determine the nucleation rates.  The crystal 
nucleation rate $k$ is related to the free energy barrier ($G(n)$) by \cite{auer_prediction_2001}
\begin{equation}
k = A e^{-\beta \Delta G(n^{*})}
\end{equation}
where
\begin{equation}
A \approx \rho f_{n^{*}} \sqrt{\frac{|\beta \Delta G^{\prime \prime} (n^{*})|}{2 \pi}},
\end{equation}
$n^{*}$ is the number of particles in the critical nucleus, 
$\rho$ is the number density of the supersaturated fluid, 
$f_{n^{*}}$ is the rate particles are attached to the critical cluster, and 
$G^{\prime \prime}$ is the second derivative of the Gibbs free energy barrier.
Auer and Frenkel \cite{auer_prediction_2001} 
showed that the attachment rate $f_{n^{*}}$ could be 
related to the mean square deviation of the cluster size at the top of the barrier by
\begin{equation}
f_{n^{*}} = \frac{1}{2} \frac{\left<\Delta n^{2}(t)\right>}{t}.
\label{eq:variance}
\end{equation}
The mean square deviation of the cluster size can then be calculated by either employing
a kinetic MC simulation or a MD simulation  at the top of the barrier.
For simplicity, in the remainder of this paper the nucleation rate determined using 
this method will be referred to as umbrella sampling (US) nucleation rates, 
although to calculate the nucleation rates both US simulations and dynamical simulations (KMC or MD)
are necessary. 
 
\begin{figure}[!hbt]
\includegraphics[width=0.45\textwidth]{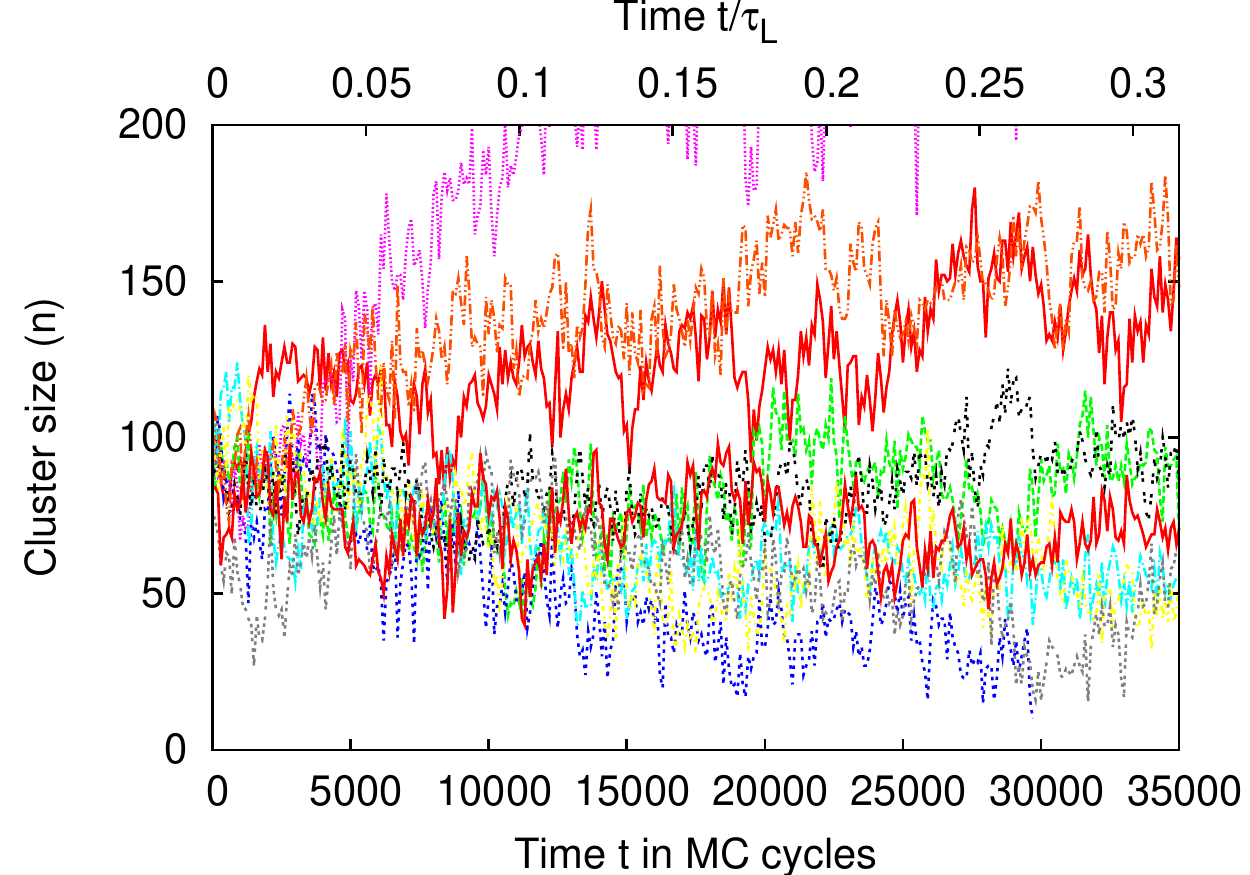}
\caption{
The cluster size ($n(t)$) as a function of time in MC cycles for a random selection of clusters that start
at the top of the nucleation barrier. \label{fig:cs}
}
\end{figure}

The mean square deviation, or variance, in the cluster size 
appearing in Eq. \ref{eq:variance}
has both a short- and long-time behaviour.
At short times, fluctuations are due to particles performing Brownian motion around
their average positions while the long-time behaviour is caused by rearrangements of particles
required for the barrier crossings. The slope of the variance is large at short times
where only the fast rattling is sampled. However, the longer the time the further the system has
diffused away from the critical cluster size at the top of the nucleation barrier.
Auer \cite{Auer_thesis} states that runs need to be selected that remain at the top of
the barrier. However, when this is done the attachment rate is lower than when the average
over all runs is taken since it excludes the runs that move off the barrier fast and have the
largest attachment rate. This problem is analogous to determining the diffusion constant 
of a particle performing a random walk. By only including walks which remain in the vicinity
of the origin, the measurement is biased and excludes trajectories which quickly 
move away from the origin. This results is an underestimation of the diffusion constant, and similarly, 
in this case, an underestimate of the attachment rate.  
In Fig.~\ref{fig:cs}
we demonstrate how, starting from a critical cluster, the size of the nucleus  fluctuates as a function of time
and, in fact, can completely disappear or double in size within $0.3/\tau_l$ where $\tau_l$ is the
time that it takes a particle on average to diffuse over a distance equal to its
diameter i.e.~$\tau_l=\sigma^2/(6D_l)$.

\begin{figure}[t]
\includegraphics[width=0.45\textwidth]{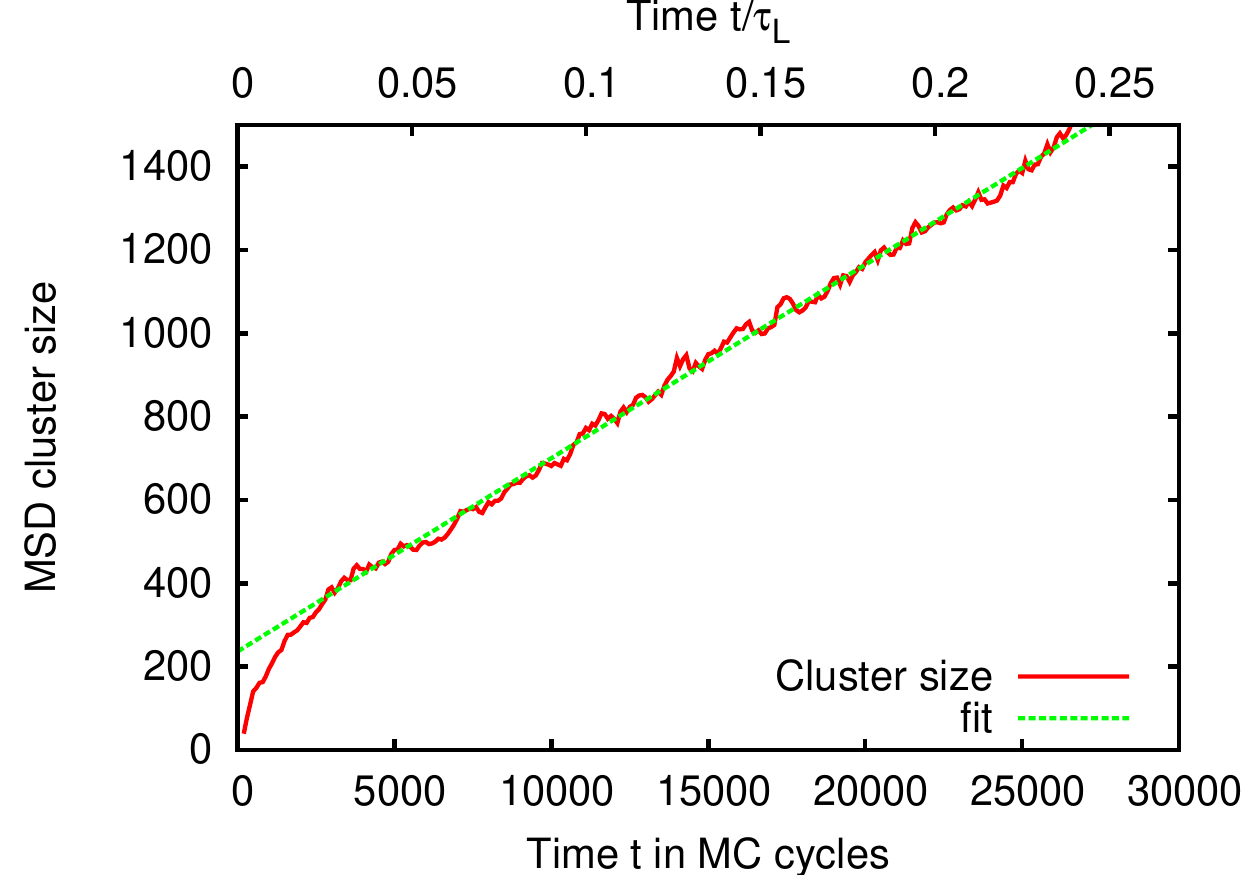}
\caption{
The mean squared deviation (MSD)  of the cluster size $\left< \Delta n^{2} (t) \right>$
as function of time $t$ in MC cycles. The cluster size has been measured
every cycle and averaged over 100 cycles to reduce the short-time fluctuations.
The slope of this graph is twice the attachment rate (Eq. \ref{eq:variance}). \label{fig:fc}
}
\end{figure}

The kinetic prefactor was determined using KMC simulations with 3000 particles
in an NVT ensemble in a cubic box with periodic boundary conditions. The initial
configurations were taken from US simulations in one of the windows at the top of the barrier.
We examined the results from both Gaussian and normally distributed Monte Carlo
steps and found agreement within the statistical errors. For all the simulations, the MC stepsize was between $0.01\sigma$  and $0.1\sigma$. 
The variance of the cluster 
size for a typical system is shown in Fig. \ref{fig:fc}. 
We observed a large variance in the rates calculated for
different nuclei. Specifically, some nuclei have attachment rates more than an order of magnitude
higher than other nuclei of similar size.
The nuclei with low attachment rates appeared to have a smoother surface than the nuclei
with a high attachment rate.

Our results for the  kinetic prefactors and 
nucleation rates for pressures $\beta p \sigma^{3}=15,16,17$  are reported in Table \ref{tab:USrates}. 

\begin{table}
\begin{tabular}{ccccccc}
\hline
\hline
$\beta p \sigma^{3}$ & $\xi_{c}$   & $n^{*}$ & $ \beta \Delta G(n^{*})$ & $\beta \Delta G''(n^{*})$ & $f_{n^{*}}/D_0$ &   $k\sigma^{5}/D_0$ \\
\hline
15 & 8   &       212     &       $42.1 \pm 0.2$ &   $-9.6 \cdot 10^{-4}$  &   661.4  &   $4.35\cdot 10^{-18}$ \\
16 & 8   &       112     &       $27.5 \pm 0.6$ &   $-1.6 \cdot 10^{-3}$  &   429.1  &   $7.80\cdot 10^{-12}$ \\
17 & 6   &       102     &       $19.6 \pm 0.3$ &   $-1.2 \cdot 10^{-3}$  &   712.9  &   $3.08\cdot 10^{-8}$ \\
17 & 8   &       72      &       $20.0 \pm 0.4$ &   $-2.0 \cdot 10^{-3}$  &   469.8  &   $1.77\cdot 10^{-8}$ \\
17 & 10  &       30      &       $19.4 \pm 0.7$ &   $-9.4 \cdot 10^{-3}$  &   316.1  &   $4.49\cdot 10^{-8}$ \\
\hline
\end{tabular}
\caption{Nucleation rates $k$ in units of $D_0/\sigma^{5}$ with $D_0$ the short time diffusion coefficient 
as a function of reduced pressure ($\beta p \sigma^{3}$) as predicted by umbrella sampling. 
$G''(n^*)$ is the second order derivative of the Gibbs free energy at the critical nucleus size $n^*$.}
\label{tab:USrates}
\end{table}


\section{Forward Flux Sampling}
\label{sec:FFS}

\subsection{Method}

The forward flux sampling method was introduced by Allen {\it et al.} \cite{allen_sampling_2005}
in 2005 to study rare events and has since been applied to a  wide variety of systems.
Two review articles (Refs. \onlinecite{allen_forward_2009,escobedo_transition_2009}) 
on the subject have appeared recently and provide a thorough 
overview of the method.  
In the present paper we discuss FFS as it 
pertains to the liquid to solid nucleation process in hard spheres.  In general, FFS follows 
the progress of a reaction coordinate during a rare event. For
hard-sphere nucleation, a reasonable reaction coordinate ($Q$) is the 
number of particles in the largest crystalline cluster in the system ($n$). 
For the remainder of this paper, for all FFS calculations, we take the reaction coordinate to be
the order parameter discussed in Sec. \ref{sec:order} with $\xi_{c}=8$, $r_c=1.3$, and $d_c=0.7 $.
In general, the reaction coordinate is used to divide 
phase space by a sequence of interfaces ({$\lambda_{0}$, $\lambda_{1}$, ... $\lambda_{N}$}) 
associated with increasing values $n(\mathbf{r}^{N})$ such that the nucleation process between any 
two interfaces can be examined. In our case the liquid is composed of all states with 
$n<\lambda_{0}$ and the solid contains all states with $n>\lambda_{N}$.
While the complete nucleation event is rare, the 
interfaces are chosen such that the part of the nucleation process between 
consecutive interfaces is not rare, and can thus be thoroughly studied. 

In the FFS methodology, the nucleation rate from the fluid phase $A$ to the solid phase $B$
is given by
\begin{eqnarray}
k_{AB}   &=& \Phi_{A \lambda_0} P({\lambda_N}|\lambda_{0}) \\
         &=& \Phi_{A \lambda_0} \prod_{i=0}^{N-1} P({\lambda_{i+1}|\lambda_{i}})
\label{eq:ffs}
\end{eqnarray}
where $\Phi_{A\lambda_0}$ is the steady-state flux of trajectories leaving the $A$ 
state and crossing the interface $\lambda_0$  in a volume $V$,
and $P({ \lambda_{i+1} | \lambda_{i}})$ is the probability that a configuration starting
at interface $\lambda_{i}$ will reach interface $\lambda_{i+1}$ 
before it returns to the fluid ($A$).

\begin{figure}[!hbt]
\includegraphics[width=0.45\textwidth]{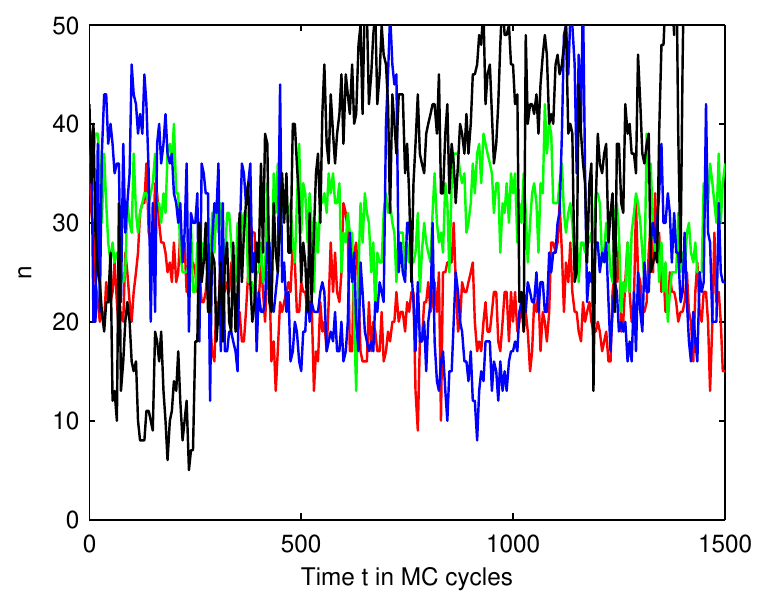}
\caption{
The cluster size as a function of time $t$ in MC cycles for 4 random trajectories at pressure $\beta p \sigma^{3}=17$
starting with a cluster size of $n=43$ using kinetic MC 
simulations with stepsize $\Delta_\mathrm{KMC}=0.1\sigma$ and measuring the order parameter every $\Delta t_\mathrm{ord} = 5$ MC steps. 
\label{fig:fluctuations} }
\end{figure}

If we apply this method directly to a hard-sphere system a number of difficulties arise.
As shown in Fig. \ref{fig:cs}, on short times the size of a cluster measured by the order parameter  
fluctuates wildly. The variance in the cluster size
displays two different types of behaviour, short-time fluctuations related to 
surface fluctuations of the cluster, and a longer time cluster growth (Fig. \ref{fig:fc}). 
Thus, if we try to measure the flux $\Phi_{A \lambda_0}$ directly, we encounter difficulties  
due to these short-time surface fluctuations. In theory, FFS should be able to handle these types of 
fluctuations, however, they increase the amount of statistics necessary to 
properly measure the flux and the first probability window properly. 
In the second part of FFS calculations, probabilities of the form $P(\lambda_{i+1}|\lambda_{i})$ need to be determined. 
In calculating these probabilities it is important to be able to determine if a cluster has returned 
to the fluid (A). 
For pre-critical clusters 
we find large fluctuations of the order parameter, 
as shown in Fig. \ref{fig:fluctuations}, which can lead to a cluster being misidentified 
as the fluid (A). Specifically, 
in this figure the darkest trajectory (black) shows a  cluster containing 43 particles 
that  shrinks to ~5 particles before it returns to 
40, and finally reaches a cluster size of 60 particles. Hence, if we had set $\lambda_0=5$, 
this trajectory would have been identified as melting back to the fluid phase (A). However,
since the growth of a cluster from size 5 to 60 is a rare event in our 
system, we presume that this was simply a short-time fluctuation of the cluster
and not a `real' melting of the instantaneously measured cluster. For pre-critical clusters, these fluctuations result
 in cluster sizes that are  smaller
than the cluster `really' is. We suggest that these fluctuations are largely related to the difficulty 
that this order parameter has in distinguishing between solid- and fluid-like particles at the fluid-solid
interface.  For larger clusters, where the surface to volume ratio is small, this problem is minimal.
However, for elongated or rough pre-critical clusters,
where the surface to volume ratio is large,
these surface fluctuations and rearrangements are important, and 
can cause problems in measuring the order parameter. 

Thus, to try and address these problems, in this paper, we apply forward flux sampling 
in a slightly novel way.  We regroup the elements of the rate calculation such that 
\begin{eqnarray}
k_{AB} &=& \tilde{\Phi}_{A \lambda_{1}} \prod_{i=1}^{N-1} P(\lambda_{i+1} | \lambda_{i}).
\label{eq:ffsnovel}
\end{eqnarray}
where 
\begin{equation}
\tilde{\Phi}_{A\lambda_1} = \Phi_{A \lambda_0}  P({\lambda_{1}|\lambda_{0}}). 
\end{equation}
We note that if $\lambda_{1}$ is chosen such it is a relatively rare event 
for trajectories starting in $A$ to reach $\lambda_{1}$, then
\begin{eqnarray}
        \tilde{\Phi}_{A\lambda_1}    \approx \frac{1}{\left<t_{A\lambda_1}\right>V}
\label{eq:ffsflux}
\end{eqnarray}
where $\left<t_{A\lambda_1}\right>$ is the average time it takes a trajectory in $A$ to reach $\lambda_{1}$.
The approximation made here, in contrast to normal FFS simulations, 
is that the time the system spends with an order parameter greater than $\lambda_{1}$ is negligible.
Since even reaching this interface is a rare event, this approximation should have a minimal effect on  
the resulting rate.
Additionally, in this way we are relatively free to place the first interface ($\lambda_{0}$) anywhere under $\lambda_{1}$.  
\footnote{While it does appear that Eq. \ref{eq:ffsnovel} is completely independent of $\lambda_0$, this is not 
strictly correct as $\lambda_0$ creates the border for state A and state A is expected to be 
a metastable, equilibrated state.  For the purposes of this paper, the difference is insignificant 
as the average time for a nucleation event is much longer than the relaxation time for the fluid.}
We choose to use $\lambda_{0}=1$ to minimize the effect of fluctuations, as seen in Fig. \ref{fig:fluctuations},  
on the probability to reach the following interface. 
Here we assume that any crystalline order in a system with an order parameter of 1 likely does not arise from fluctuation of a 
much larger cluster, but rather is very close to the fluid, and is expected to 
fully melt and not grow out to the next interface.
In this manner we are able to start several parallel trajectories from the fluid 
in order to measure $\left<t_{A\lambda_1}\right>$, stopping whenever the trajectory first hits interface $\lambda_{1}$.

%

In our implementation of FFS, we employ kinetic Monte Carlo (KMC) simulations at fixed pressure 
to follow the trajectories from the liquid to the solid.
The KMC simulations are characterized by two parameters, the maximum stepsize ($\Delta_\mathrm{KMC}$)
per attempt to move each particle, and the frequency with which the order parameter (reaction 
coordinate) is measured $\Delta t_\mathrm{ord}$.
However, during an FFS simulation, it is expected that the 
order parameter is known at all times such that it is possible 
to identify exactly when and if a given simulation reaches 
an interface.  Thus it is possible that $\Delta t_\mathrm{ord}$ 
introduces an additional error into our measurement of the rate.

To examine the effects of i) the approximation associated with our method for 
calculating $\tilde{\Phi}_{A\lambda_1}$,
ii) the short-time fluctuations of the order parameter (which could be considered as an error in the 
measurement of the cluster size), and iii) the frequency of measuring the order parameter,  
 we examined the nucleation rate for a 
simple one-dimensional model system in the presence of such features. 
Details of these simulations are given 
in Appendix A.  In this simple model system, we find that none of these features
have a large effect on the rate.  In fact, for most cases,
the difference is too small to see within our error bars.



\begin{table*}
\begin{tabular}{l l l l l l l l l l l l l}
\hline
\hline
$\Delta_\mathrm{KMC}$   &0.1	&0.1&	0.1&	0.2&	0.2&	0.2&	0.2&	0.2&	0.2&	0.2&	0.2&	0.2\\
$\Delta t_\mathrm{ord}$       &2	&2&	2&	2&	2&	2&	1&	1&	1&	10&	10&	10\\
\hline
$P(\lambda_{2}|\lambda_{1})$ &0.112	&0.103	&0.139	&0.101	&0.105	&0.132	&0.112	&0.146	&0.138	&0.122	&0.127	&0.146\\
$P(\lambda_{3}|\lambda_{2})$ &0.096	&0.117	&0.090	&0.104	&0.093	&0.112	&0.115	&0.097	&0.079	&0.103	&0.081	&0.080\\
$P(\lambda_{4}|\lambda_{3})$ &0.128	&0.117	&0.074	&0.116	&0.111	&0.161	&0.151	&0.110	&0.110	&0.121	&0.091	&0.116\\
$P(\lambda_{5}|\lambda_{4})$ &0.180	&0.159	&0.082	&0.156	&0.115	&0.241	&0.209	&0.189	&0.173	&0.121	&0.073	&0.150\\
$P(\lambda_{6}|\lambda_{5})$ &0.167	&0.154	&0.149	&0.225	&0.148	&0.256	&0.274	&0.151	&0.189	&0.189	&0.121	&0.187\\
$P(\lambda_{7}|\lambda_{6})$ &0.071	&0.074	&0.060	&0.128	&0.093	&0.118	&0.121	&0.052	&0.092	&0.169	&0.077	&0.064\\
$P(\lambda_{8}|\lambda_{7})$ &0.104	&0.078	&0.051	&0.109	&0.091	&0.109	&0.119	&0.077	&0.126	&0.132	&0.087	&0.064\\
$P(\lambda_{9}|\lambda_{8})$ &0.100	&0.100	&0.105	&0.083	&0.075	&0.089	&0.101	&0.081	&0.129	&0.101	&0.109	&0.068\\
\hline
$P(\lambda_{9}|\lambda_{1})$    &$3\cdot 10^{-8}$ & $2\cdot 10^{-8}$ & $4\cdot 10^{-9}$ & $5\cdot 10^{-8}$ & $1\cdot 10^{-8}$ & $2\cdot 10^{-7}$ &  $2\cdot 10^{-7}$ & $1 \cdot 10^{-8}$ & $ 6\cdot 10^{-8}$ & $8 \cdot 10^{-8}$ & 	$6 \cdot 10^{-9}$ & $1 \cdot 10^{-8}$ \\
\hline
\end{tabular}
\caption{Probabilities $P(\lambda_{i+1}|\lambda_i)$ for the first 8 interfaces for a pressure of $\beta p \sigma^{3}=15$ where the 
KMC  simulations stepsize ($\Delta_\mathrm{KMC}$) and  the number of MC steps between measuring the order parameter $\Delta t_\mathrm{ord}$ are varied.
The following interfaces were used: $\lambda_{2}=20$, $\lambda_{3}=26$, 
$\lambda_{4}=32$, $\lambda_{5}=38$, $\lambda_{6}=44$, $\lambda_{7}=54$, $\lambda_{8}=65$, and  $\lambda_{9}=78$.  In all cases, 
100 configurations were started in the fluid and reached the first interface, and at each interface, $C_{i}=10$ copies of each
successful configuration were used.
\label{tab:FFStests}
}
\end{table*}

\begin{table}
\begin{tabular}{l l l l l}
\hline
\hline
$\beta p \sigma^{3}$ & $\lambda_{1}$ & $\tilde{\Phi}_{A\lambda_1}/6D_l$ & $P(\lambda_B | \lambda_1)$ & $R/6D_l$ \\
\hline
17                   & 27            &   $2.66 \cdot 10^{-5}$                            &   $7.6 \cdot 10^{-3}$      &   $2.0\cdot 10^{-7}$      \\
17                   & 27            &   $2.68 \cdot 10^{-5}$                            &   $1.4 \cdot 10^{-2}$      &   $3.7\cdot 10^{-7}$      \\
16                   & 20            &   $8.57 \cdot 10^{-6}$                            &   $3.1 \cdot 10^{-7}$      &   $2.6\cdot 10^{-12}$     \\
16                   & 20            &   $8.57 \cdot 10^{-6}$                            &   $2.1 \cdot 10^{-7}$      &   $1.8\cdot 10^{-12}$     \\
15                   & 15            &   $8.72 \cdot 10^{-6}$                            &   $1.9 \cdot 10^{-15}$     &   $1.6\cdot 10^{-20}$     \\
\hline
\end{tabular}
\caption{Nucleation rates predicted using forward flux sampling in short-time diffusion coefficient  units ($D_{0}$). 
The probabilities $P(\lambda_B | \lambda_1)$, number of steps between the order parameter measurements $\Delta_{\mathrm{ord}}$, and 
kinetic MC stepsize are as in Tables  \ref{tab:FFS17A},  \ref{tab:FFS16A}, and \ref{tab:FFS15A}. At each
interface, $C_{i}$ copied of each successful configuration were used.
\label{tab:FFSrates}
}
\end{table}

\begin{table}
\begin{tabular}{l l | l l | l l}
\hline
\hline
 &             & trial 1 & & trial 2  & \\
i& $\lambda_i$ & $C_{i-1}$ & $P(\lambda_{i}|\lambda_{i-1})$ & $C_{i-1}$ & $P(\lambda_{i}|\lambda_{i-1})$ \\
\hline
2 & 43 & 10  &0.137 &10 &0.157 \\ 
3 & 60 & 10  &0.272 &10 &0.312\\
4 & 90 & 10  &0.350 &10 &0.414\\
5 & 150& 2   &0.594 &2  &0.691\\
6 & 250& 2   &0.988 &2  &0.988\\
\hline
\end{tabular}
\caption{Probabilities $P(\lambda_{i+1}|\lambda_i)$ for the interfaces used in calculating the nucleation rate for 
pressure $\beta p \sigma^{3}=17$ with step size 
$\Delta_\mathrm{KMC}=0.1\sigma$ and  measuring the order parameter every $\Delta t_\mathrm{ord}=5$ MC cycles.
\label{tab:FFS17A}
}
\end{table}

\begin{table}
\begin{tabular}{l l | l l | l l}
\hline
\hline
 &             & trial 1 & & trial 2  & \\
i& $\lambda_i$ & $C_{i-1}$ & $P(\lambda_{i}|\lambda_{i-1})$ & $C_{i-1}$ & $P(\lambda_{i}|\lambda_{i-1})$ \\
\hline
2 &28   &10 &0.105 &10 &0.110\\
3 &38   &10 &0.075 &10 &0.077\\
4 &50   &10 &0.070 &10 &0.089\\
5 &70   &10 &0.114 &10 &0.089\\
6 &90   &10 &0.095 &10 &0.101\\
7 &110  &10 &0.339 &10 &0.278\\
8 &250  &10 &0.152 &10 &0.112\\
9 &350  &1 &1.000 &1 &1.000\\
\hline
\end{tabular}
\caption{Same as Table \ref{tab:FFS17A} but for 
 $\beta p \sigma^3 = 16$.
\label{tab:FFS16A}
}
\end{table}

\begin{table}
\begin{tabular}{l l | l l}
\hline
\hline
i& $\lambda_i$ & $C_{i-1}$ & $P(\lambda_{i}|\lambda_{i-1})$ \\
\hline
2& 20  &10 &0.101 \\
3& 26  &10 &0.104 \\
4& 32  &10 &0.116 \\
5& 38  &10 &0.156 \\
6& 44  &10 &0.225 \\ 
7& 54  &10 &0.128\\ 
8& 65  &10 &0.109\\ 
9& 78  &10 &0.083\\
10& 92  &10 &0.101\\
11& 110 &10 &0.085\\
12& 135 &10 &0.062\\
13& 160 &10 &0.131\\
14& 190 &10 &0.131\\
15& 230 &10 &0.134\\
16& 400  &10 &0.058\\
\hline
\end{tabular}
\caption{Same as Table \ref{tab:FFS17A} but for $\beta p \sigma^3 = 15$  and with  $\Delta t_\mathrm{ord}=2$.
\label{tab:FFS15A}
}
\end{table}

\subsection{Simulation details and results}
All simulations were performed with 3000 particle in a cubic
box with periodic boundary conditions.
Initial configurations were produced using NPT MC simulations of a liquid phase  
at a reduced pressure of $\beta p \sigma^3 = 1000$.
The simulations were stopped when the packing fraction associated with the pressure of
interest was reached. This initial configuration was then relaxed using an NPT simulation
at the correct pressure ($\beta p \sigma^3 = 15,16,17$).  The relaxation 
consisted of at least 10,000 MC cycles, after which the simulation continued until    
a measurement of the  order parameter found no crystalline particles in the system.

In order to determine the flux and the probabilities, $100$ trajectories were started in the liquid and 
terminated when $n(\mathbf{r}^{N})=\lambda_{1}$. These trajectories were produced using  KMC simulations.
The probability $P(\lambda_2 | \lambda_1)$
was then found by making  $C_{1}$ copies of the configurations that  reached $\lambda_{1}$, and following
these configurations until they either reached $\lambda_{2}$ or returned to the fluid. By taking different random 
number seeds, the various copies of the same configurations follow 
different trajectories. The fraction of successful trajectories corresponds to the required probability.
The successful trajectories  were then copied $C_{2}$ times to determine $P(\lambda_3|\lambda_2)$.
The remaining $P(\lambda_{i+1}| \lambda_i)$'s are calculated similarly. 
%

To study the effect of the two KMC parameters, namely $\Delta_\mathrm{KMC}$ and $\Delta t_\mathrm{ord}$,
on the nucleation rates, we have examined the first 8 FFS windows 
for $\beta p \sigma^{3}=15$ for various values of the number of MC steps between the order parameter measurements $\Delta t_\mathrm{ord}$ and 
the maximum displacement $\Delta_\mathrm{KMC}$ for the KMC simulations.
The results are shown in Table \ref{tab:FFStests}.  As shown in this table we do not find a significant effect on the rate
from either parameter.  Thus for numerical efficiency, unless otherwise indicated, the
rates in this section come from  $\Delta t_\mathrm{ord} = 5$ MC cycles and 
 $\Delta_\mathrm{KMC} = 0.2\sigma$.

For pressures $\beta p \sigma^{3}=16$ and 17 we have performed two separate FFS calculations to determine 
the nucleation rates, and for pressure $\beta p \sigma^{3}=15$ we have the result from a single FFS simulation.
A summary of the results are given in Table \ref{tab:FFSrates}. A complete summary of the results for
$P(\lambda_i+1|\lambda_i)$ 
for each simulation is given in Tables 
 \ref{tab:FFS17A},  \ref{tab:FFS16A}, 
 and \ref{tab:FFS15A}.


\section{Summary and Discussion}
\label{sec:results}

\subsection{Nucleation Rates}

\begin{figure*}
\centering
\includegraphics[width=0.65\textwidth]{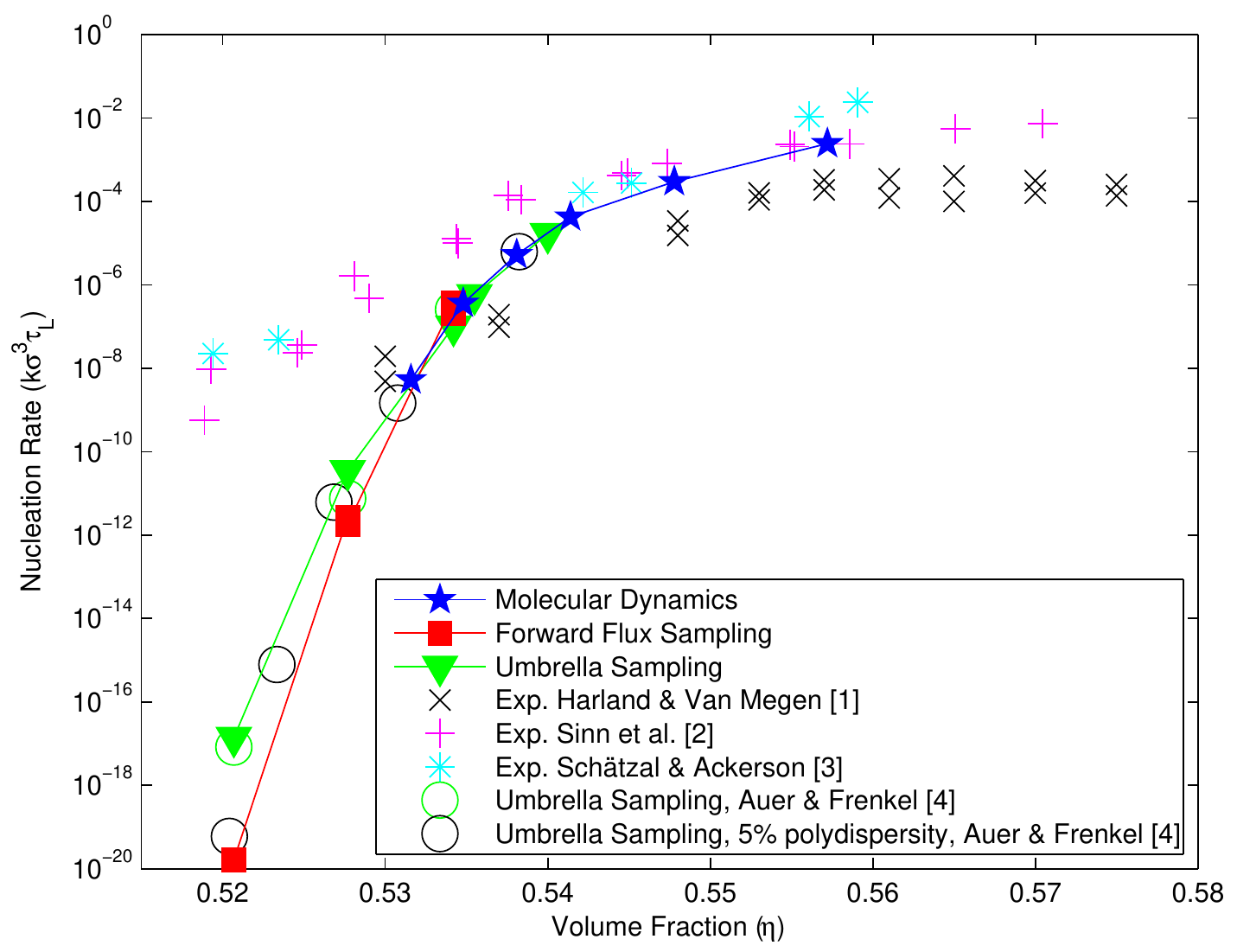}
\label{fig:rates}
\caption{\label{fig:rates}  A comparison of the crystal nucleation rates  of hard spheres as determined by the three methods described in this 
paper FFS, US, and MD with the experimental results from Refs. \onlinecite{harland_crystallization_1997,sinn_solidification_2001,schatzel_density_1993} and 
previous theoretical results from Ref. \onlinecite{auer_prediction_2001}. Note that error bars have not been included in this plot.
In general, the error bars of the simulated nucleation rates are largest for lower supersaturations (ie. lower volume fractions), as the
barrier height is higher.  For the FFS and US simulations, the error for $\beta p \sigma^{3}=15$ ($\eta=0.5214$) 
is between 2 and 3 orders of magnitude,
and for $\beta p \sigma^{3}=17$ ($\eta=0.5352$) is approximately one to two orders of magnitude.  
The MD results are quite accurate around $\beta p \sigma^{3}=17$, however
the error bars are larger for the higher pressure MD results. Within these estimated error bars, the simulated nucleation rates are all in 
agreement, while the experimentally obtained rates show a markedly different behaviour, particularly for low supersaturations where the 
the difference between the simulations and experiments can be as large as 12 orders of magnitude. }
\end{figure*}

In this section we examine hard-sphere nucleation rates predicted using  US simulations, MD simulations
and FFS simulations together with the experimental results of 
Harland and Van Megen, \cite{harland_crystallization_1997}
Sinn {\it et al.} \cite{sinn_solidification_2001} and 
Sch\"atzel and Ackerson \cite{schatzel_density_1993}
 and the US simulations of monodisperse 
and 5\% polydisperse hard-spheres mixtures examined by Auer and Frenkel. \cite{auer_prediction_2001}  
The experimental volume fractions have been scaled to yield the
coexistence densities of monodisperse hard spheres. \cite{pusey_hard_2009}
Similarly, we scale the polydisperse results of Auer and Frenkel with the 
coexistence densities determined in Ref. \onlinecite{fasolo_fractionation_2004}.
Inspired by the recent work of Pusey {\it et al.}, \cite{pusey_hard_2009} we
plot the nucleation rates in units of the long-time diffusion coefficient.  
In experiments with colloidal particles, the influence of the solvent on the 
dynamics cannot be ignored. Specifically, the system slows down due to hydrodynamic interactions 
when the density is increased.  
However, since hydrodynamics are included in the long-time diffusion units,
if we present the nucleation rates in
terms of the long-time diffusion coefficient, our predicted nucleation rates should be in agreement with
the experiments.  
The time in experiments is typically measured in units of $D_0$, the free diffusion at low
density. We convert the short-time diffusion coefficient $D_0$ to  long-time diffusion coefficient $D_l$ 
using 
\begin{equation}
\frac{D_L (\eta)}{D_0} = \left( 1-\frac{\eta}{0.58} \right)^{\delta}.
\end{equation}
Harland and Van Megen \cite{harland_crystallization_1997}  claim that $\delta=2.6$ gives a good fit to their system
and Sinn {\it et al.} \cite{sinn_solidification_2001} use $\delta=2.58$. Since the system Sch\"atzel and  Ackerson \cite{schatzel_density_1993} 
examine is very similar to the other two, we use $\delta=2.6$ to convert their nucleation rates to long-time units.  
We note that both $\delta=2.58$ and $\delta=2.6$ give very similar results.
The results for both the theoretical and experimental rates in long time units are shown in Fig. \ref{fig:rates}.

In Ref. \onlinecite{pusey_hard_2009}, Pusey {\it et al.} showed that the nucleation rates for 
various polydispersies (0 to 6\%) of hard-sphere mixtures
collapsed onto the same curve when the rates were plotted in units of the long-time diffusion coefficient.
We find similar results here. Both the monodisperse and polydisperse 
US results of Auer and Frenkel, \cite{auer_prediction_2001} in addition to our own US
predictions of the nucleation rate, agree well within the expected measurement error.
Additionally, we find that the simulation results of the US, FFS, and MD all 
agree.  

However, on the experimental side, the nucleation rates of  Harland and Van Megen\cite{harland_crystallization_1997}
are approximately  one to two orders of magnitude below the experiments of Sinn {\it et al.} \cite{sinn_solidification_2001} 
and Sch\"atzel and  Ackerson. \cite{schatzel_density_1993} This
is unexpected due to the similarity between the experimental systems. In our opinion, 
the main difference between these experiments is the polydispersity of the particle mixtures: 
5\% in the case of Harland and Van Megen,\cite{harland_crystallization_1997}
2.5\% in the case of Sinn {\it et al.}, \cite{sinn_solidification_2001} and $<5$\% 
for Sch\"atzel and  Ackerson. \cite{schatzel_density_1993}
However, as demonstrated by Pusey {\it et al.}, \cite{pusey_hard_2009} and now also in Fig. \ref{fig:rates},
the nucleation rate when measured in long-time diffusion coefficient units should not be effected by the polydispersity. Thus, this
seems unlikely as an explanation. A more probable difference between the results may simply be due to measurement error
in the experimental volume fraction which is extremely difficult to measure.  
As shown in Fig. \ref{fig:rates}, a measurement error of $\Delta \eta=0.05$ can have a
very large effect on the nucleation rates.
It is worth mentioning here that the units of Ref. \onlinecite{harland_crystallization_1997} were not always
mentioned, and thus there also may be some error in the manner in which we converted these rates 
to long-time units. 

When we compare the experimental rates with the theoretical results, we find 
that while the experiments appear to match the general trend of the simulations for high supersaturations
they predict a significantly  higher nucleation rate at lower densities.
The argument presented above regarding measurement error in the 
volume fraction is one possible explanation, ie. by simply shifting the experimentally
 predicted nucleation rates to higher densities the agreement is much better. However, 
we speculate that there is another possible reason for the discrepancy.  Specifically, at high 
supersaturations there should be many nucleation events occurring in the experimental
system during a fairly short time interval. Hence, it should be possible to measure the 
nucleation rate before a single cluster has the chance to grow out significantly.  However, at lower
supersaturations, when a nucleation event is extremely rare, a single cluster in the 
experimental system can grow out significantly before sufficient nucleation events have occurred 
to measure the rate.  However, these large clusters can contain a number of twinning defects, \cite{snook}
 resulting in  scattering from the various crystalline domains. 
Scattering from these domains may lead to an over-count in the 
nucleation events per volume and time unit,
yielding higher nucleation rates. 

\begin{figure*}
\begin{tabular}{c c c}
$\xi_c=5$ & $\xi_c=7$ & $\xi_c=9$ \\
\includegraphics[width=0.32\textwidth]{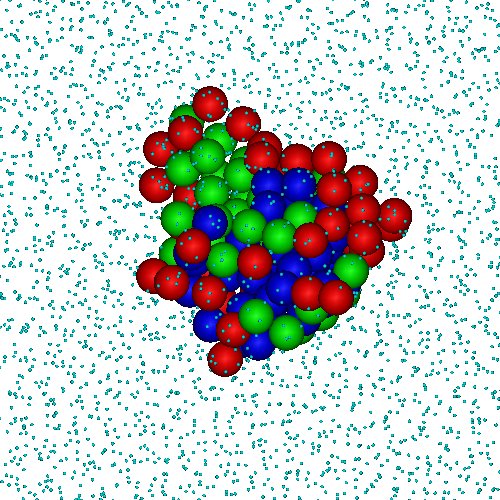}&
\includegraphics[width=0.32\textwidth]{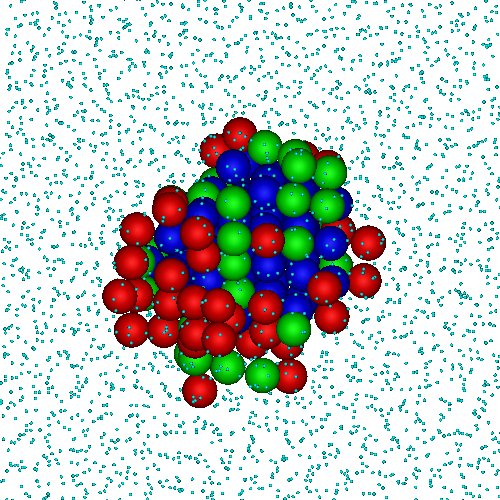}&
\includegraphics[width=0.32\textwidth]{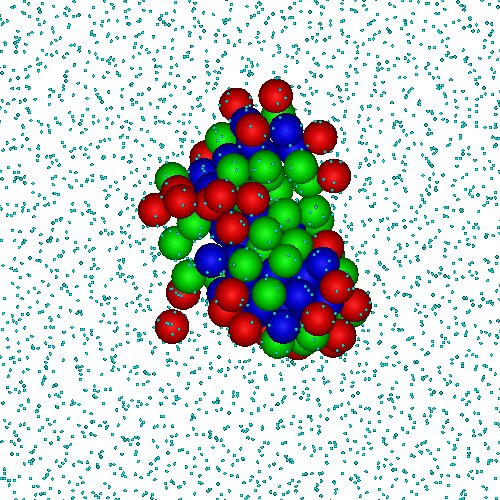}\\
\includegraphics[width=0.32\textwidth]{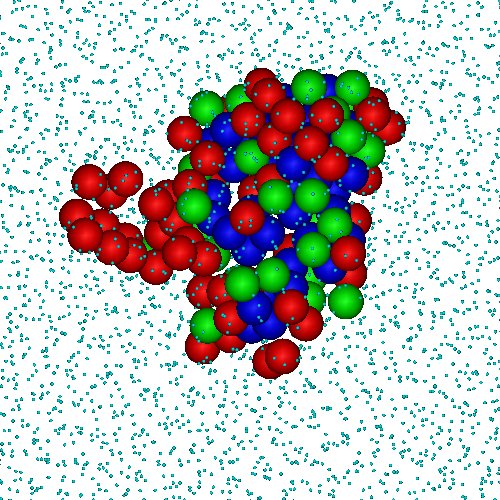}&
\includegraphics[width=0.32\textwidth]{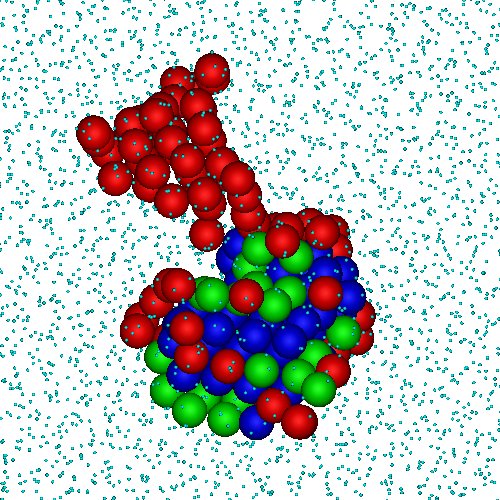}&
\includegraphics[width=0.32\textwidth]{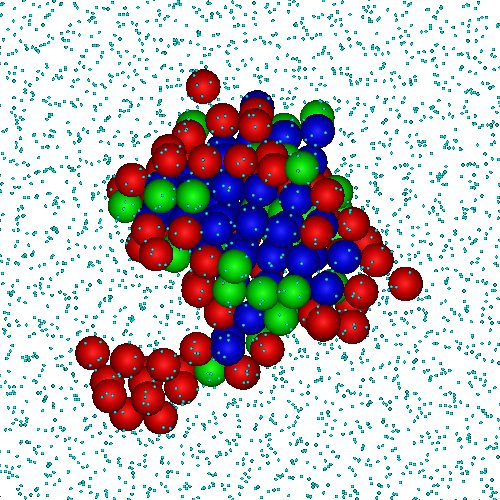}\\
\end{tabular}
\caption{
Two typical snapshots (top and bottom) of the critical nuclei as obtained with US at a volume fraction $\eta=0.5355$ using
different values of the critical number of crystalline bonds $\xi_c=5$ (left), 7 (middle) and 9 (right) in the biasing potential.
The clusters are analyzed with three different crystalline order parameters.
The blue particles are found by all three cluster criteria, the green particles have $\xi=7$ or
8 crystalline bonds and the red particles have only $\xi=5$ or 6 crystalline bonds.
\label{fig_ccsnap}
}
\end{figure*}

\begin{figure*}
\begin{tabular}{c c c}
\includegraphics[width=0.32\textwidth]{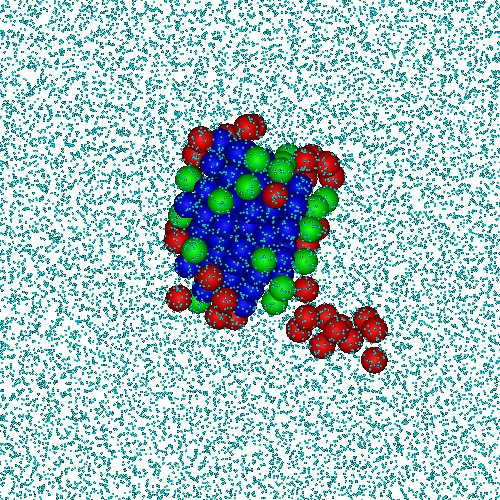}&
\includegraphics[width=0.32\textwidth]{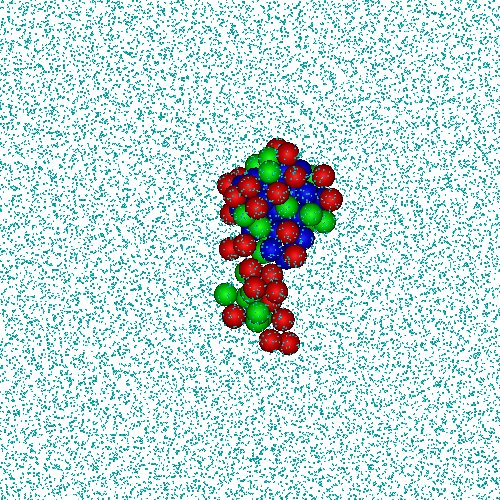}&
\includegraphics[width=0.32\textwidth]{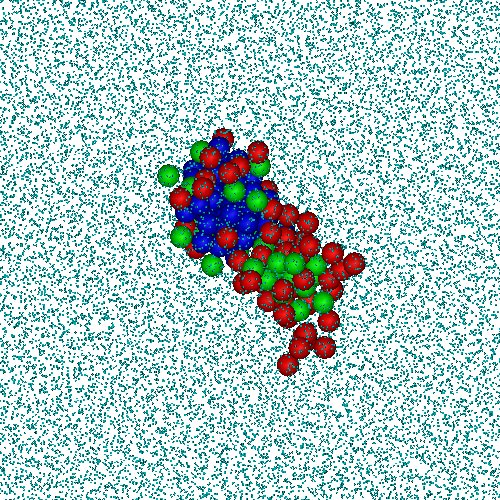}\\
\end{tabular}
\caption{
Snapshots of spontaneously formed nuclei during an MD simulation at a volume fraction of $\eta=0.537$.
The snapshots were taken just before the nuclei grew. The color coding of the particles is the same as in Fig.~\ref{fig_ccsnap}.
\label{fig_ccmd}
}
\end{figure*}

\subsection{Nuclei}

To examine whether the structure and shape of the critical clusters from US simulations depended on
the precise threshold values used for the crystalline order parameters, we compared and analysed
the critical clusters obtained when three different crystalline order parameters were used 
 to bias the US simulations, namely, $\xi_{c}=5,7$ and 9. Subsequently we analyzed these critical clusters
using the three different order parameters.
In Fig.~\ref{fig_ccsnap},  two typical critical clusters from different  biasing order 
parameters are shown on the top and bottom rows.   
The nucleus of the cluster, shown in blue,
was identified by all three cluster criteria ($\xi_{c}=5,7$ and 9). The main difference between the criteria is the
location of the fluid-solid interface as shown by the green and red particles. The strictest order parameter
finds only the more ordered center whereas the loosest version detects the more disordered
particles at the interface as well.

If Fig.~\ref{fig_ccmd} we show some of the nuclei obtained from MD simulations. These snapshots
were taken just before the nuclei grew out so they are not necessarily precisely at the top
of the nucleation barrier. They appear very similar in roughness and aspect ratio to those
obtained from US simulations.

\begin{figure}
\centering
\includegraphics[width=0.45\textwidth]{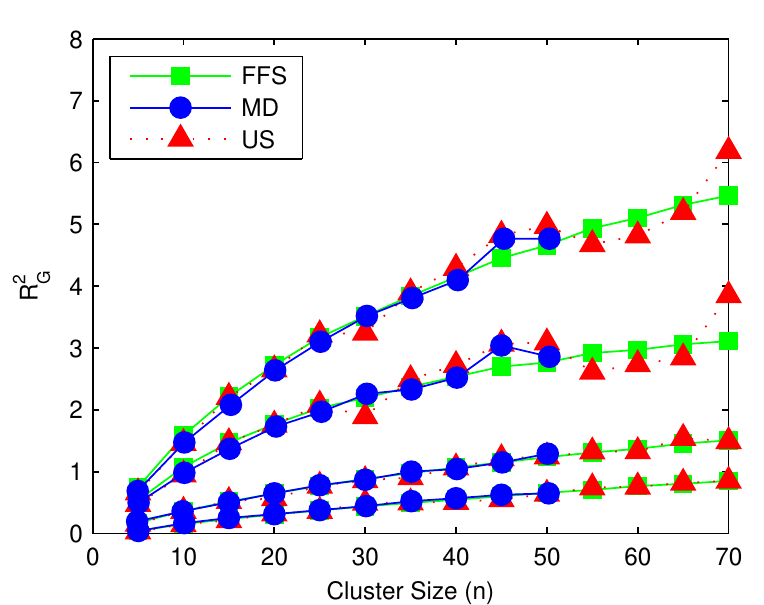} 
\caption{\label{fig:clusters}  A comparison of the three components of the
radius of gyration tensor as a function of cluster size $n$, as well as the sum of the three components, 
for clusters produced using  FFS, MD, and US simulations.
}
\end{figure}

To further examine whether
the choice of method influenced the resulting clusters, we calculated the radius of gyration tensor 
for each of the methods 
for pressure $\beta p \sigma^3=17$ as a function of cluster size (see Figure \ref{fig:clusters}).
There is no indication that the clusters in any of the simulation methods differed substantially.

\begin{figure}
\centering
\includegraphics[width=0.45\textwidth]{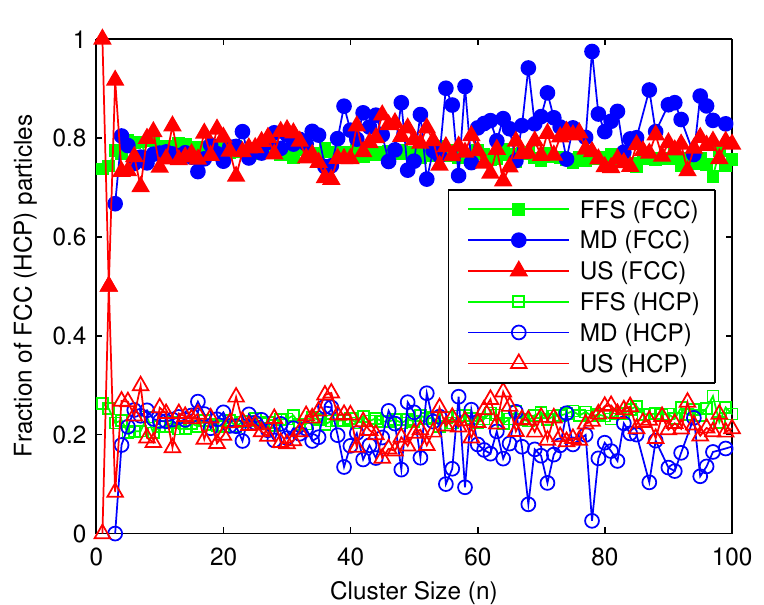} 
\caption{\label{fig:FCCHCP} Fraction of particles identified as either FCC or HCP respectively in the
clusters produced via molecular dynamics (MD), forward flux sampling (FFS), and umbrella sampling (US) simulations
as a function of cluster size $n$.  All three methods agree and find the pre-critial 
clusters prodominately FCC. 
}
\end{figure}

Additionally, we examined whether the simulation technique influenced the type of pre-critical 
nuclei that formed in the simulations, ie. face-centered-cubic (FCC), and hexagonal-close-packed (HCP).
To do this we used the order parameter introduced by Ref. \onlinecite{lechner_accurate_2008} which allows us to 
identify each particle in the cluster as either FCC-like or HCP-like.   
The results for a wide range in nucleus size is shown in Fig. \ref{fig:FCCHCP}.  
We find complete agreement between the three simulation techniques.  Specifically, in all cases we find
that the nucleus is composed of approximately 80\% FCC-like particles.  This was unexpected 
as the free energy difference between the bulk FCC and HCP phases is about 0.001k$_{B}$T per 
particle at melting \cite{bolhuis_entropy_1997} and hence 
a random stacking of hexagonal layers in the nuclei would be expected. \cite{pronk_can_1999}
We speculate that this predominance of FCC stacking in the nuclei arises from surface effects.

\section{Conclusions}
\label{sec:conclusions}

In conclusion, we have examined crystal nucleation of hard spheres with 
molecular dynamics, umbrella sampling and forward flux sampling simulations.
We find that the nucleation rates predicted by  all three methods agree over the large range in volume
fractions we examined.  Additionally, in agreement with the recent 
work of Pusey {\it et al.}, \cite{pusey_hard_2009} we find that by measuring the nucleation rates 
in terms of the long-time diffusion constant and scaling to the coexistence 
density of monodisperse hard spheres, the 5\% polydisperse results 
of Auer and Frenkel  \cite{auer_prediction_2001}  also agree. On examining the critical clusters, we do not find 
a difference in the nuclei formed using the three simulation techniques.

We have also compared our nucleation rates with previous experimental data, 
specifically, the nucleation rates predicted by Harland and Van Megen, \cite{harland_crystallization_1997} 
Sinn {\it et al.} \cite{sinn_solidification_2001} and Sch\"atzel and Ackerson. \cite{schatzel_density_1993}
The  nucleation rates measured by these three experiments, in contrast to what would be expected, 
differ by about one order of magnitude.  
In general, the experimental systems
are similar enough that one would have expected agreement in the rate once the 
rates were scaled to the coexistence densities of hard spheres. 
Additionally, while the simulation results 
agree well with the experimental results for high supersaturations, 
there is a significant difference between the simulations and experiments for smaller 
volume fractions.  We speculate here that this difference may be due to difficulties 
in distinguishing between separate nuclei domains in the experiments, or measurement error
in the experimental volume fractions.


\section{Acknowledgements}
We would like to thank Frank Smallenburg, Matthieu Marechal, Eduardo Sanz and Chantal Valeriani for many useful discussions.  
We acknowledge financial
support from the NWO-VICI grant and the high potential programme from Utrecht University.


\appendix

\section{FFS in the presence of measurement error}

\begin{table*}
\begin{tabular}{cccccc}
\hline
\hline
$\Delta t_{\mathrm{ord}}$ &                1   &                      2 &                   5    &                     10 &                50  \\
\hline
	 & $1.2723\cdot 10^{-12}$ &  $ 1.0589\cdot 10^{-12}$ & $  1.8075\cdot 10^{-12}$ & $  1.5455\cdot 10^{-12}$ & $  1.3835\cdot 10^{-12}$ \\
	 & $1.3780\cdot 10^{-12}$ &  $ 1.7217\cdot 10^{-12}$ & $  1.3314\cdot 10^{-12}$ & $  1.4461\cdot 10^{-12}$ & $  1.0666\cdot 10^{-12}$ \\
	 & $1.2364\cdot 10^{-12}$ &  $ 1.2924\cdot 10^{-12}$ & $  1.4847\cdot 10^{-12}$ & $  1.1482\cdot 10^{-12}$ & $  1.6134\cdot 10^{-12}$ \\
	 & $1.6942\cdot 10^{-12}$ &  $ 1.6422\cdot 10^{-12}$ & $  1.9482\cdot 10^{-12}$ & $  1.4383\cdot 10^{-12}$ & $  1.7550\cdot 10^{-12}$ \\
	 & $1.2662\cdot 10^{-12}$ &  $ 1.2340\cdot 10^{-12}$ & $  1.5692\cdot 10^{-12}$ & $  1.6060\cdot 10^{-12}$ & $  1.2908\cdot 10^{-12}$ \\
	 & $1.6918\cdot 10^{-12}$ &  $ 1.3530\cdot 10^{-12}$ & $  1.6238\cdot 10^{-12}$ & $  1.6244\cdot 10^{-12}$ & $  1.4012\cdot 10^{-12}$ \\
	 & $1.4646\cdot 10^{-12}$ &  $ 1.1788\cdot 10^{-12}$ & $  1.6928\cdot 10^{-12}$ & $  1.0191\cdot 10^{-12}$ & $  1.3403\cdot 10^{-12}$ \\
	 & $1.6809\cdot 10^{-12}$ &  $ 1.5860\cdot 10^{-12}$ & $  1.1903\cdot 10^{-12}$ & $  1.6227\cdot 10^{-12}$ & $  1.0582\cdot 10^{-12}$ \\
	 & $1.4602\cdot 10^{-12}$ &  $ 1.7018\cdot 10^{-12}$ & $  1.3191\cdot 10^{-12}$ & $  1.3850\cdot 10^{-12}$ & $  2.3732\cdot 10^{-12}$ \\
	 & $1.7459\cdot 10^{-12}$ &  $ 1.9154\cdot 10^{-12}$ & $  1.5638\cdot 10^{-12}$ & $  1.2378\cdot 10^{-12}$ & $  1.2692\cdot 10^{-12}$ \\
\hline
Avg. Rate & $1.5\cdot 10^{-12}$ &  $ 1.5\cdot 10^{-12}$ & $  1.6\cdot 10^{-12}$ & $  1.4\cdot 10^{-12}$ & $  1.5\cdot 10^{-12}$ \\
Std. Error & $6.0\cdot 10^{-14}$ &  $ 8.4\cdot 10^{-14}$ & $  7.0\cdot 10^{-14}$ & $  6.3\cdot 10^{-14}$ & $  1.2\cdot 10^{-13}$ \\
\hline
\end{tabular}
\caption{Nucleation rates for the one-dimensional potential given by Eq. \ref{eq:toypotential} and shown in Fig. \ref{fig:toypotential} 
for $\Delta t_{\mathrm{ord}}$  as indicated. 
For each $\Delta t_{\mathrm{ord}}$ ,
we performed 10 independent FFS simulations.  The average rate and associated standard deviation  is also as indicated.
In all cases, 100 configurations were started in the fluid, and at each interface $C_{i}=10$ copies of the successful configurations 
were used to calculate the proceeding probabilities.  The interfaces were placed at $\lambda_{0}=0$, 
$\lambda_{1}=1.5$, $\lambda_{2} = 1.7$, $\lambda_{3} = 1.9$, $\lambda_{4}=2.2$, $\lambda_{5}=2.6$, $\lambda_{6}=3.3$, and $\lambda_{7}=4.0$ and 
the flux was calculated using Eq. \ref{eq:ffsflux}.}
\label{tab:FFStoy}
\end{table*}

\begin{table*}
\begin{tabular}{cccccc}
\hline
\hline
$\sigma_{\mathrm{Gauss}}$          &  0.02    &     0.04 &    0.06    &   0.08 &   0.1  \\
\hline
&$1.8623\cdot 10^{-12}$   &      $   1.7281\cdot 10^{-12}$        & $ 1.2630\cdot 10^{-12}$&     $   1.0634\cdot 10^{-12}$       &  $ 1.9158\cdot 10^{-12}$ \\
&$1.7627\cdot 10^{-12}$   &      $   1.6090\cdot 10^{-12}$        & $ 1.6402\cdot 10^{-12}$&     $   1.5655\cdot 10^{-12}$       &  $ 1.8785\cdot 10^{-12}$ \\
&$9.9796\cdot 10^{-13}$   &      $   1.6305\cdot 10^{-12}$        & $ 1.5799\cdot 10^{-12}$&     $   1.6936\cdot 10^{-12}$       &  $ 1.4937\cdot 10^{-12}$ \\
&$1.3743\cdot 10^{-12}$   &      $   1.2261\cdot 10^{-12}$        & $ 1.8305\cdot 10^{-12}$&     $   1.7733\cdot 10^{-12}$       &  $ 1.1142\cdot 10^{-12}$ \\
&$1.6917\cdot 10^{-12}$   &      $   1.8054\cdot 10^{-12}$        & $ 1.6191\cdot 10^{-12}$&     $   1.8941\cdot 10^{-12}$       &  $ 1.0402\cdot 10^{-12}$ \\
&$1.1842\cdot 10^{-12}$   &      $   1.3337\cdot 10^{-12}$        & $ 1.3283\cdot 10^{-12}$&     $   1.4039\cdot 10^{-12}$       &  $ 7.0735\cdot 10^{-13}$ \\
&$1.5289\cdot 10^{-12}$   &      $   8.6859\cdot 10^{-13}$        & $ 1.3129\cdot 10^{-12}$&     $   2.7115\cdot 10^{-12}$       &  $ 2.4711\cdot 10^{-12}$ \\
&$1.8918\cdot 10^{-12}$   &      $   1.4325\cdot 10^{-12}$        & $ 1.3203\cdot 10^{-12}$&     $   1.3792\cdot 10^{-12}$       &  $ 1.6288\cdot 10^{-12}$ \\
&$1.3144\cdot 10^{-12}$   &      $   1.2283\cdot 10^{-12}$        & $ 1.0459\cdot 10^{-12}$&     $   1.7194\cdot 10^{-12}$       &  $ 1.3764\cdot 10^{-12}$ \\
&$1.6654\cdot 10^{-12}$   &      $   1.1236\cdot 10^{-12}$        & $ 1.2572\cdot 10^{-12}$&     $   1.9631\cdot 10^{-12}$       &  $ 1.8976\cdot 10^{-12}$ \\
\hline
Avg. Rate  &$1.5\cdot 10^{-12}$      &      $   1.4\cdot 10^{-12}$           & $ 1.4\cdot 10^{-12}$&        $   1.7\cdot 10^{-12}$          &  $ 1.6\cdot 10^{-12}$ \\
Std. Error&$9.5\cdot 10^{-14}$       &      $   9.4\cdot 10^{-14}$           & $ 7.5\cdot 10^{-14}$&        $   1.4\cdot 10^{-13}$          &  $ 1.6\cdot 10^{-13}$ \\
\hline
\end{tabular}
\caption{Nucleation rates for the one-dimensional potential given by Eq. \ref{eq:toypotential} and shown in Fig. \ref{fig:toypotential} 
where the order parameter is given by Eq. \ref{eq:noise} and $\sigma_{\mathrm{Gauss}}$ is as indicated.   
For each $\sigma_{\mathrm{Gauss}}$,
we performed 10 independent FFS simulations.  The average rate and associated standard deviation  is also as indicated.
In all cases, 100 configurations were started in the fluid, and at each interface $C_{i}=10$ copies of the successful configurations 
were used to calculate the proceeding probabilities.  The interfaces were placed at $\lambda_{0}=0$, 
$\lambda_{1}=1.5$, $\lambda_{2} = 1.7$, $\lambda_{3} = 1.9$, $\lambda_{4}=2.2$, $\lambda_{5}=2.6$, $\lambda_{6}=3.3$, and $\lambda_{7}=4.0$ and 
the flux was calculated using Eq. \ref{eq:ffsflux}.}
\label{tab:FFStoy2}
\end{table*}

As mentioned in Section \ref{sec:FFS} of this paper, 
the FFS technique assumes that the reaction coordinate is known exactly at all times.
However, for the hard-sphere system examined in this paper,  this is not possible due to the 
computational time required for measuring the order parameter.
In applying the FFS technique to hard spheres,  
two separate types of error are introduced: 
i)
error associated with our inability to know the value of the reaction coordinate at
all times, and 
 ii) 
 an error in measuring the 
number of particles in a cluster for a given configuration.
Additionally, as discussed in Section \ref{sec:FFS}, in this paper we have
applied FFS in a slightly novel manner.
In this appendix, we introduce a simple model to examine the effect this approximation and the effect
such measurement errors have on the nucleation rate predicted by 
forward flux sampling.

\begin{figure}[t!]
\centering
\includegraphics[width=0.45\textwidth]{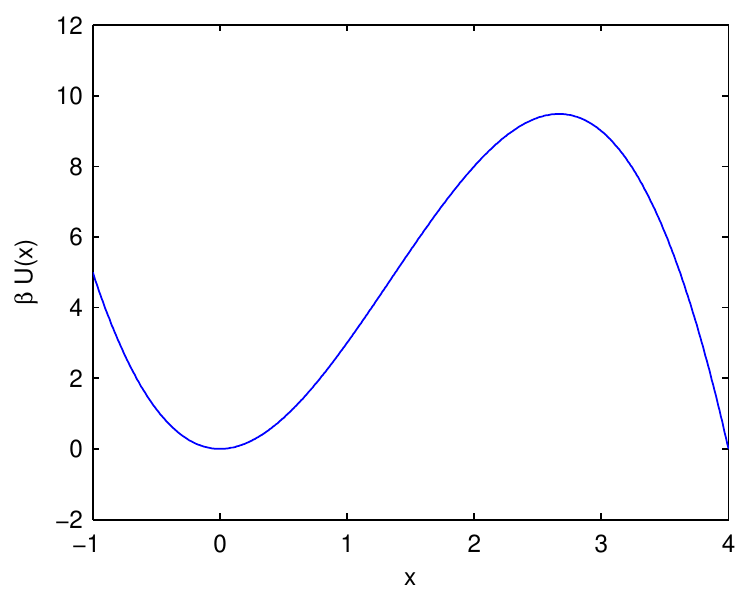}
\label{fig:toypotential}
\caption{\label{fig:toypotential} Toy model potential used to study forward flux sampling in the
present of various types of measurement error.
}
\end{figure}

To this end, we study the transition rate for a single Brownian particle to
surmount a one dimensional potential energy barrier given by
\begin{equation}
\beta U(x) = 8x^{2} - 2x^3.
\label{eq:toypotential}
\end{equation}
A plot of the barrier is shown in Fig. \ref{fig:toypotential}.  For this potential, 
we consider the `liquid' state to be near $x=0$ and the `solid' phase to be near $x=4$.

We first determine the `exact' nucleation rate using spontaneous simulations.  
To do this we perform a random walk starting at $x=0$ and determine the time 
it takes the random walk to surmount the barrier.  The rate is then given by $R=1/\left<t\right>$.
Performing 40 such random walks we find the nucleation rate to be $1.5\cdot 10^{-12}$.   
In all the calculations in this section, we set the KMC stepsize equal to $\Delta_{KMC} =0.1$.

Secondly we explore the effect on the nucleation rate of not knowing the value of the order parameter at all times.
For this purpose we have performed FFS simulations when the order parameter was measured every $\Delta t_{\mathrm{ord}}=1,2,5,10,50$
kinetic Monte Carlo steps.  The results are shown in Table \ref{tab:FFStoy}.  The average nucleation rates predicted for 
all values of $\Delta t_{\mathrm{ord}}$ clearly are the same within error.  Similarly, the 
standard error associated with $\Delta t_{\mathrm{ord}}=1,2,5,10$ are approximately the same, and is only marginally 
larger for $\Delta t_{\mathrm{ord}}=50$.  Hence we conclude that the frequency of measuring the order parameter 
does not significantly affect the predicted nucleation rate.  Additionally, these nucleation rates agree with the
nucleation rate predicted from spontaneous simulations indicating that of applying FFS as outlined
in Section \ref{sec:FFS} predicts the correct nucleation rates.

Finally, we examine the effect that the measurement error in the cluster size has on the nucleation rate.  
For this purpose, we apply a noise term to our order parameter such that 
\begin{equation}
x_{m} = x_{\mathrm{true}}+\delta
\label{eq:noise}
\end{equation}
where $x_{m}$ is the value of the order parameter used in the FFS simulation, $x_{\mathrm{true}}$ is the 
true value of the order parameter, and  $\delta$ is taken from a Gaussian distribution with a mean of 
0 and a standard deviation $\sigma_{\mathrm{Gauss}}$.
In Table \ref{tab:FFStoy2} we demonstrate the effect on the predicted nucleation rate for 
various choices of $\sigma_{\mathrm{Gauss}}$.  The resulting nucleation rates are in good 
agreement with the spontaneous results.  For larger $\sigma_{\mathrm{Gauss}}$, eg. $\sigma_{\mathrm{Gauss}}=0.08$ and 0.1,
the standard error in the results is slightly larger, however, the predicted nucleation rates
are still correct.

In summary, we have examined the effect of the approximation described by Eq. \ref{eq:ffsflux}, as well 
as the effect of measurement error in the order parameter and the measurement frequency $\Delta t_{\mathrm{ord}}$
of the order parameter. We do not find a significant effect on the predicted nucleation rates.  
Thus we conclude that FFS should be robust to the types of error we are introducing 
when we apply the technique to hard spheres.


\end{document}